\documentclass[12pt]{article}
\usepackage[agsm, abbr]{harvard}
\usepackage{graphicx}
\usepackage{amsmath}
\usepackage{pdfpages}
\usepackage{graphics}
\usepackage{xypic}
\usepackage[cm]{fullpage}
\usepackage{enumerate}
\usepackage{bbm}
\usepackage{graphicx}
\usepackage{graphics}
\usepackage{fancybox}
\usepackage{amsmath,float,latexsym,psfrag,epsf,epsfig,amssymb,
rotating}
\usepackage{color}
\usepackage{url}
\usepackage{fullpage}
\usepackage[ruled,linesnumbered]{algorithm2e}
\usepackage{algorithmicx}
\usepackage{caption}
\usepackage{subcaption}
\newcommand{\x}{{\mathbf{x}}}
\newcommand{\bX}{{\mathbf{X}}}

\newcommand{\bY}{{\mathbf{Y}}}

\newcommand{\X}{{\mathbf{X}}}
\newcommand{\Y}{{\mathbf{Y}}}

\newcommand{\bc}{{\mathbf{c}}}

\newcommand{\I}{{\mathbf{I}}}
\newcommand{\diff}{\mathrm{d}}

\newcommand{\bftheta}{\hbox{\boldmath$\theta$}}

\newcommand{\bfomega}{\hbox{\boldmath$\omega$}}

\newcommand{\bfmu}{\hbox{\boldmath$\mu$}}

\title{ Bayesian model selection for the latent position cluster model for Social Networks }

\author{Caitr\'{i}ona Ryan$^*$, Jason Wyse$^{\dagger}$ and Nial Friel$^{\ddagger}$ 
\\
{\small $^*$ Department of Mathematics and Statistics, University of Limerick,}\\{\small Limerick, Ireland (e-mail: \texttt{Caitriona.Ryan@ul.ie})}\\
{\small $^{\dagger}$ Discipline of Statistics, School of Computer Science and Statistics, Trinity College Dublin,}\\{\small College Green, Dublin 2, Ireland (e-mail: \texttt{wyseja@tcd.ie})}\\
{\small $^{\ddagger}$ School of Mathematics and Statistics and Insight: The National Centre for Big Data Analytics,}\\
{\small  University College Dublin, Belfield, Dublin 4, Ireland (e-mail: \texttt{nial.friel@ucd.ie})}}

\begin{document}

 \maketitle

\begin{abstract}

The latent position cluster model is a popular model for the statistical analysis of network data. This model assumes that 
there is an underlying latent space in which the actors follow a finite mixture distribution. Moreover, actors which are close in 
this latent space are more likely to be tied by an edge. This is an appealing approach since it allows the model to cluster actors which 
consequently provides the practitioner with useful qualitative information. However, exploring the uncertainty in the number 
of underlying latent components in the mixture distribution is a complex task. The current state-of-the-art is to use an
approximate form 
of BIC for this purpose, where an approximation of the log-likelihood is used instead of the true log-likelihood which is unavailable. 
The main contribution of this paper is to show that through the use of conjugate prior distributions it is possible to analytically 
integrate out almost all of the model parameters, leaving a posterior distribution which depends on the allocation vector of the mixture model. This enables posterior inference over the number of components in the latent mixture
distribution without using trans-dimensional MCMC algorithms such as reversible jump MCMC. Our approach is compared with the state-of-the-art \texttt{latentnet} \cite{Kriv:Hand13} and \texttt{VBLPCM}~\cite{salter:murphy12} packages.

\end{abstract}

\noindent { \textbf{Key words:}  collapsed latent position cluster model; reversible jump Markov chain Monte Carlo; Bayesian model choice; 
social network analysis; finite mixture model}

\section{Introduction}

A social network consists of nodes or actors in a graph, for example, individuals or organizations, 
connected by one or more specific types of interdependency, such as, friendship, business relationships or trade between 
countries.  
The analysis of network data has a rich interdisciplinary history finding application in a wide range of areas including 
sociology \cite{wasserman:galaskiewicz1994}, 
physics \cite{adamicetal01}, 
biology \cite{michailidis12}, 
computer science \cite{faloutsosetal1999}
and many more. The aims of network analysis are both descriptive and inferential.
For example, one might be interested in examining global structure within a network or in analysing network attributes such 
as the degree distribution as well as the local structure such as the identification of influential or highly connected actors 
in the network. Inferential goals include hypothesis testing, model comparison and making predictions, for example, how far
will a virus spread through a network.

There have been many statistical models proposed for the analysis of network data, the most popular of which include the 
exponential random graph model see \citeasnoun{wasserman:pattison1996} and \citeasnoun{robinsetal07b}
 and the stochastic block model of \citeasnoun{nowicki:snijders01} and its variants. 
For a recent perspective on the statistical analysis of network data, see \citeasnoun{kolaczyk09}.
An alternative and popular approach to modelling network data is the latent space approach \cite{hoff:raft:hand02}. 

Here each actor is embedded in a latent `social space' in which actors that are close in the latent space are more likely to be tied by an edge. 
Latent space models naturally accommodate many sociological features such as homophily, reciprocity 
and transitivity.
The recent development of \citeasnoun{handcocketal07} extends the latent space model of \citeasnoun{hoff:raft:hand02}
 to cluster actors directly, where the positions of actors are assumed to be distributed according
to a finite mixture. 
The latent position cluster model (LPCM) provides a useful interpretation of the network since the underlying latent model 
provides an automatic means of clustering actors while also providing the uncertainty around the probability of actor membership
to each cluster. The R package \texttt{latentnet} \cite{Kriv:Hand07,Kriv:Hand13}, which is part of the \texttt{statnet} 
suite of packages, can be used to fit an LPCM.

Despite its popularity, a major difficulty with LPCMs is inferring the number of components in the 
latent finite mixture distribution. The approach advocated by \citeasnoun{handcocketal07} is to assess this uncertainty by estimating the Bayesian
information criterion (BIC) for each possible model. However, it turns out that it is computationally prohibitive to calculate the maximum log-likelihood
used in BIC. A tractable approximation is to condition on the minimum Kullback-Leibler  estimate of the actors latent positions \cite{shortreedetal06},
rather than integrating over the posterior distribution of the actors positions and accounting for the uncertainty in
these latent positions. Note that a variational Bayes approximation has been proposed by \citeasnoun{salter:murphy12} implemented in the R package \texttt{VBLPCM}, but it too uses the same strategy as \citeasnoun{handcocketal07} to infer the number of components.
One of the primary contributions of this article is to resolve this issue. To this end we use conjugate prior distributions which
allow almost all latent mixture parameters to be integrated out. This results in a collapsed posterior distribution which depends on the
vector of allocations of actors to components. The important consequence of this is that the allocation vector encodes the number of components of the mixture distribution, but crucially,
the number of components can be inferred without the use of 
trans-dimensional MCMC techniques such as reversible jump MCMC \cite{richardson:green97}. This approach is similar 
to that presented in \citeasnoun{nob:fearn07} and \citeasnoun{wyse:friel12} for the collapsed finite mixture model and
latent block models, respectively. 

The software which
accompanies this paper can be used to implement all the examples presented herein.

The paper begins in Section \ref{lpcm} by describing the LPCM and the current approach to inferring the number of clusters.
Section \ref{collapsing} introduces the collapsed form of the model. 
Cross-model inference is described for the collapsed LPCM in Section~\ref{collapsing}. 
Section~\ref{sec:examples} applies and compares the methodology to current methods for some known social network data. We carry out a simulation study in Section~\ref{sec:simulated}.
Some discussion follows in Section~\ref{discussion}.
 
\section {Latent Position Cluster Model}\label{lpcm}

\subsection{Motivation and notation}

Let $\bY$ denote an observed $n \times n$ adjacency matrix indicating the presence or absence of ties between a set of $n$ actors, with $y_{ij}$ indicating presence or absence of a tie between $i$ and $j$. 
The latent position cluster model \cite{handcocketal07} and the preceeding latent position model \cite{hoff:raft:hand02} (which is a special case of the model in~\citeasnoun{handcocketal07}) assume that each actor has a corresponding position $\x$ in a latent space, usually $\mathbb{R}^d$ with  $d=2$. 

The probability of a dyadic link (or edge) is modelled using the distance between actors' positions in the latent space. Specifically, the linear predictor
\begin{equation}
\eta_{ij} = \beta - || \x_i - \x_j ||  \label{eq:linpred}
\end{equation}
gives the probability of a link between actors $i$ and $j$ through a logistic link function
\[
\Pr \left( Y_{ij} = 1 | \x_i, \x_j, \beta \right) = \frac{1}{1 + \mathrm{e}^{-\eta_{ij}}}.
\]
The appearance of the euclidean norm $|| \cdot ||$ in (\ref{eq:linpred}) measuring the latent distance between actors $i$ and $j$ has an appealing intuitive interpretation; actors who are farther apart in latent space are less likely to be tied. The parameter $\beta$ is often referred to as the abundance; high values of $\beta$ imply a high probability of forming ties (hence abundant).

A {\it local independence} assumption is made which assumes dyadic links arise independently over pairs of actors in the network. The likelihood of observing the adjacency $\bY$ then factors as a product over outcomes for dyads $\mathcal{D}$:
\begin{equation}
p(\bY\,|\,\bX,\beta) = \prod_{(i,j)\in \mathcal{D} } \Pr\left( Y_{ij} = y_{ij} | \x_i, \x_j, \beta \right)  \label{eq:like}
\end{equation}
where $\bX$ is used to collectively denote the joint positions $\x_1,\dots,\x_n$. Self ties are not allowed, meaning that for directed networks
\[
\mathcal{D} = \left\{\, (i,j) : 1 \leq i,\,j \leq n,\,  i\ne j\, \right\},
\]
while for undirected networks
\[
\mathcal{D} = \left\{ \,(i,j) : 1 \leq i \leq n, j < i\, \right\}.
\]

Below the data level, one assumes a prior on the latent positions $\bX$. \citeasnoun{hoff:raft:hand02} assumed a spherical $d$-dimensional Gaussian scaled by a precision (hyper)parameter independently for each $\x_i$. \citeasnoun{handcocketal07} extend this construction by assuming a finite mixture of $G$ $d$-dimensional Gaussians with spherical precision in place of the single Gaussian, giving joint prior on the latent positions
\begin{equation}
\pi(\bX|\bftheta,G) = \prod_{i=1}^n \left( \sum_{g=1}^G \omega_g \,\mathcal{N}( \,\x_i \,;\, \bfmu_g, 1/\tau_g \I  ) \right). \label{eq:jmix}
\end{equation}
The parameter $\bftheta$ will be taken to denote the mixture weights $\omega_g$ (which sum to one: $\sum_{g=1}^G \omega_g = 1$), the component centres and precisions $\bfmu_g, \tau_g, \, g = 1,\dots,G$. Clustering in the network can be captured by clustering in the latent positions; different clusters are represented by the components of the finite mixture. One introduces labels $\bc = (c_1,\dots,c_n)$, denoting the component to which each actor belongs. Using the labels, the joint prior density of the latent positions and labels is 
\begin{equation}
\pi(\bX,\bc | \bftheta, G ) = \prod_{g=1}^G \,\prod_{i\, : \,c_i = g} \omega_g \, \mathcal{N}(\x_i ;  \bfmu_g, 1/\tau_g \I) \label{eq:mixprior}
\end{equation}

\subsection{Bayesian LPCM} \label{sec:priors}

The Bayesian LPCM assumes priors on the LPCM parameters $\bftheta$. Independent priors are assumed for the component weights (Dirichlet), centres (Gaussian) and precisions (gamma) over the $G$ groups:
\[
\begin{array}{l}
(\omega_1,\dots,\omega_g) \sim  \mathcal{D}( \alpha, \dots,\alpha)\\
\\
\bfmu_g |\tau_g \sim  \mathcal{N}(\mathbf{0},  1/(\kappa \tau_g) \I) \qquad g=1,\dots, G \\
\\
\tau_g  \sim \mathcal{G} ( \delta/2 , \gamma/2 )\qquad g=1,\dots,G
\end{array}
\]
where $\alpha, \kappa, \delta$ and $\gamma$ are parameters to be chosen.     Choosing $\alpha=3$, $\delta = 2$ and $\gamma = 0.103$  corresponds to the prior choices made in ~\citeasnoun{handcocketal07} 
(their parameters are denoted $\nu=3$, $\alpha=2$ and $\sigma_0^2=0.103$, respectively). Our specification of the prior precision on the cluster means $\bfmu_g$ is different. We scale the within cluster precision $\tau_g$ by a factor $\kappa$. We note that values of $\kappa$ less than 1 imply that the cluster means are more dispersed than the cluster members. The prior assumed for the intercept parameter $\beta$ in the linear predictor (\ref{eq:linpred}) is $\mathcal{N}(0,2)$ as in~\citeasnoun{handcocketal07}.

The main motivation of this paper is to explore the uncertainty in $G$, the number of finite mixture components grouping the latent positions. The primary justification of the LPCM is that one interprets components in 
latent space as substantive clusters in the network. Thus, the value of $G$ is of great importance. Different values of $G$ can lead to different observations on the global properties of the network. Search strategies 
for comparing values of $G$ are discussed later, however, now the dependence of the model's core structure on the value  of $G$ is made explicit, and a prior $\pi(G)$ is assumed for $G$. \citeasnoun{nobile07} gives a 
convincing argument to take a Poisson distribution with rate 1 for $\pi(G)$. \citeasnoun{miller15} reaffirm the argument of \citeasnoun{nobile07} in their detailed discussion of eliciting priors for the number of components. 
The posterior of the LPCM, including uncertainty for $G$, may be written hierarchically as 
\begin{equation}
\pi(\bX,\bc,\bftheta,G|\Y) \propto p(\Y|\bX,\beta)\, \pi( \beta ) \, \pi( \bX,\bc | \bftheta,  G )   \, \pi( \bftheta | G )\,  \pi( G ). \label{eq:model}
\end{equation}

\section{Marginalized model approach}\label{collapsing}

An innovative and appealing characteristic of our proposed approach is that uncertainty in the latent actors' positions as well as the structural components of their behaviour (i.e. the number, $G$, of components in the  finite mixture) can be explored jointly and in tandem. Before introducing our novel approach, Section~\ref{bicsec} describes existing ``best practice'' for choosing the number of  components $G$. Then in Section~\ref{sec:cprior} and subsequent Sections, we describe the model and estimation techniques we propose.

\subsection{Existing approaches for choosing the number of components $G$} \label{bicsec}

The model marginal likelihood or model evidence~\cite{friel:wyse12} is used for model comparison in the Bayesian paradigm. For the model (\ref{eq:model}) above, the ``model'' refers to the number of components $G$ in the finite mixture, considered with the network likelihood based on the latent positions. The marginal likelihood is 
\begin{eqnarray}
\pi(\Y|G)& = & \int \int \int p(\Y|\bX,\beta)\, \pi( \beta ) \, p( \bX | \bftheta,  G )   \, \pi( \bftheta | G ) \,\diff \beta  \,\diff\bftheta \,\diff \bX \nonumber \\
  & = &\int \left[ \int  p(\Y|\bX,\beta)\, \pi( \beta )  \, \diff \beta \right] \left[\int \, \pi( \bX | \bftheta,  G ) \, \pi( \bftheta | G ) \,\diff\bftheta \right]  \,\diff \bX \label{eq:ml}
\end{eqnarray}
It is used to compute Bayes Factors and posterior model probabilities when a collection of candidate models are considered. The posterior probability of $G$ components can be evaluated via $\pi(G|\Y) \propto \pi(\Y|G)\pi(G)$. Note here the use of the {\it incomplete} joint mixture density $p( \bX | \bftheta,  G )$, (\ref{eq:jmix}). The functional form of the joint densities involved in the marginalization (\ref{eq:ml}) make it intractable, and thus approximations must be used, either simulation based, or approximations based on point estimates.

The Bayesian Information Criterion (BIC) ~\cite{schwarz1978} approximates the negative of twice the log  of the marginal likelihood. For the LPCM, a pragmatic approach to choosing the value of $G$ adopted by \citeasnoun{handcocketal07} is to condition on a fixed estimate of latent actor locations $\widehat\X = \left\{ \widehat{\x}_1,\dots,\widehat{\x}_n\right\}$ and use
\begin{equation}
-2 \log\pi(\Y|G) \approx -2 \log \left[\int p(\Y|\widehat{\bX},\beta)\, \pi( \beta ) \, \diff \beta\right] - 2 \log  \left[ \int \pi( \widehat{\bX} | \bftheta,  G )   \, \pi( \bftheta | G ) \,\diff\bftheta \right] ,\label{eq:bic}
\end{equation}
with the smallest value giving the ``best'' $G$. This can be seen as a term for a logistic regression on the network dyad values plus a term for the latent mixture model. As the logistic regression model is a function of distances between actors rather than the actual latent positions, these actor locations $\widehat{\X}$ are found by minimizing the Kullback-Leibler divergence between the true unknown model distances and the MCMC sample position based distances (see Appendix A of \citeasnoun{handcocketal07} for further details). The BIC for $G$ components then approximates the right hand side of (\ref{eq:bic}), which can be viewed as the BIC of a logistic regression ($\mathrm{BIC}_{\mathrm{LR}}$) on the observed links in the network plus the BIC of a finite mixture of spherical Gaussians with $G$ components ($\mathrm{BIC}_{\mathrm{MIX}}$). Conditioning on a $\widehat{\bX}$, \citeasnoun{handcocketal07} propose using
\begin{eqnarray*}
\mathrm{BIC}_{\mathrm{LR}} & = & -2 \log p(\Y|\widehat{\bX},\widehat{\beta}(\widehat{\bX}) ) + d_{\mathrm{LR}} \log n_{\mathrm{LR}};  \\
\mathrm{BIC}_{\mathrm{MIX}} & = & -2 \log \pi(\widehat{\bX} |\widehat{\bftheta}(\widehat{\bX})) +  d_{\mathrm{MIX}}  \log n.
\end{eqnarray*}
Here, the estimates $\widehat{\beta}$ and $\widehat{\bftheta}$ are found conditional on the positions $\widehat{\bX}$. 

There are a number of different values for the effective sample size $n_{\mathrm{LR}}$ that could be used in computing the BIC. The current version of \texttt{latentnet}~\cite{Kriv:Hand13} uses the number of links (or edges) in the network as default. Alternative choices are the number of dyads ($n(n-1)/2$ for an undirected network) or the number of actors in the network. The effective sample size used for the mixture BIC is the number of actors in the network (i.e. a unit of information for each latent position). In the case of the mixture, one could question whether an effective sample size of $n$ is a good choice for a prior on latent unobserved data.

We also note that the approximation relies entirely on one modal value of $\bX$. If the posterior of $\bX$ is highly peaked, with small uncertainty, this could appear a good approximation at face value. However, our experience is that the posterior of $\bX$ can exhibit varying degrees of spread. This plug-in approach also comes with the obvious caveat that if the ``modal'' plug-in configuration $\bX$ is suboptimal, then there could be further error introduced into the approximation of $\log \pi(\bY|G)$ which is difficult to quantify. Both \texttt{latentnet} and \texttt{VBLPCM} approximate the BIC using this plug-in approach. Envisaging and quantifying the sources and magnitude of error are open problems with potentially many factors influencing the quality of these approximations. Our proposed approach, outlined in the next section explores the posterior uncertainty in $G$ in a principled and efficient manner.

\subsection{Finite mixture prior on latent positions and marginalized posterior}\label{sec:cprior}

Following the choice of (hyper)priors on the finite mixture model parameters $\bftheta$ in Section~\ref{sec:priors}, it is possible to marginalize these parameters from the model and work with a marginalized posterior in order to search over the joint space of $(\beta, \bX, G, \bc)$. The full posterior is 
\begin{eqnarray*}
\pi(\bX,\bc,\bftheta,G|\Y) &\propto& p(\Y|\bX,\beta)\, \pi( \beta ) \, \pi( \bX,\bc | \bftheta,  G )   \, \pi( \bftheta | G )\,  \pi( G ) \\
&\propto& p(\Y|\bX,\beta)\, \pi( \beta ) \, \pi( \bX,\bc, \bftheta,  G ).
\end{eqnarray*} 
The joint density $\pi( \bX,\bc, \bftheta,  G )$ is explicitly
\begin{eqnarray*}
\pi( \bX,\bc, \bftheta,  G ) & \propto & \pi(G)\, \times \,\frac{\Gamma(G\alpha)}{\Gamma(\alpha)^G} \prod_{g=1}^G \omega_g^{\alpha-1}\,\times \,\prod_{g=1}^G\pi(\tau_g) \pi(\bfmu_g|\tau_g) \,\prod_{i\, : \,c_i = g} \omega_g \, \mathcal{N}(\x_i ;  \bfmu_g, 1/\tau_g \I) \\
& = & \pi(G) \,\times\, \frac{\Gamma(G\alpha)}{\Gamma(\alpha)^G} \prod_{g=1}^G \omega_g^{n_g + \alpha-1}\,\times\, \prod_{g=1}^G \pi(\tau_g) \pi(\bfmu_g|\tau_g) \prod_{i:c_i=g} \mathcal{N}( \x_i; \bfmu_{g}, 1/\tau_{g} \mathbf{I} ).
\end{eqnarray*}
The marginalized posterior is obtained by marginalizing the elements of $\bftheta$:
\begin{eqnarray*}
\pi(\bX,G , \bc  ) &\propto & \pi( G) \,\times\, \frac{\Gamma(G\alpha)}{\Gamma(\alpha)^G} \int \prod_{g=1}^G \omega_g^{n_g + \alpha-1} \,\diff \bfomega  \,\times\, \prod_{g=1}^G \int\left[ \pi(\tau_g) \pi(\bfmu_g|\tau_g) \prod_{i:c_i=g} \mathcal{N}( \x_i; \bfmu_{g}, 1/\tau_{g} \mathbf{I} ) \right]\, \diff \bfmu_g \, \diff \tau_g .\\
\end{eqnarray*}
Following this, one may write
\begin{eqnarray*}
\pi(\bX,G , \bc  ) &\propto & \pi(G)\, \pi(\bc | G ) \, \pi(\bX | \bc,  G ),
\end{eqnarray*}
where 
\[
\pi(\bc | G ) = \frac{\Gamma( G \alpha )}{\Gamma(\alpha)^G }\, \frac{\prod_{g=1}^G \Gamma(n_g + \alpha)}{\Gamma(n + G \alpha)}
\]
and 
\begin{eqnarray*}
\pi(\bX| \bc,  G ) & = & \prod_{g=1}^G  \int \pi(\tau_g) \pi(\bfmu_g| \tau_g) \prod_{i:c_i = g} \mathcal{N}(\x_i; \bfmu_g,  \tau_g \I ) \,\diff \bfmu_g \,\diff \tau_g  \\
& = & \prod_{g=1}^G \pi^{-n_g d / 2}\frac{\gamma^{\delta/2}}{ (n_g / \kappa + 1) ^{d/2}} \frac{\Gamma\left((n_g d + \delta)/2\right)}{\Gamma\left(\delta/2\right)}  \left[\, \sum_{i:c_i=g} || \x_i ||^2 - \frac{|| \sum_{i:c_i = g} \x_i ||^2}{n_g + \kappa}  + \gamma\, \right]^{-(n_g d + \delta)/2}\\
& =& \prod_{g=1}^G \lambda_g( \bX,\bc ),
\end{eqnarray*}
where $n_g = \#\{i : c_i =g\}$ and $\lambda_g(\bX,\bc)$ denotes the joint component marginal likelihood for observations in group $g$. 

Writing the marginalized model in full as
\begin{equation}
\pi(\bX,G,\bc,\beta|\Y) \propto  p(\Y|\bX,\beta)\, \pi( \beta ) \, \, \pi( \bX, G, \bc ) \label{eq:post}
\end{equation}
makes explicit the structure of the marginalized posterior. Now the joint prior on $(\bX,G,\bc)$ is continuous in $\bX$ but discrete in $\bc$. Thus, stochastic searches over a discrete space can be used to search over finite mixtures with different numbers of components and allocations, in order to obtain samples from the posterior defined by the right hand side of (\ref{eq:post}). Thus, reversible jump steps \cite{richardson:green97} may be avoided when searching over candidate mixture models. This is beneficial, both from the point of view of having a reduced parameter space, as well as avoiding the difficult task of proposing between model moves. This approach has been used successfully by~\citeasnoun{nob:fearn07} for Gaussian finite mixtures and~\citeasnoun{wyse:friel12} for model-based bi-clustering. A key difference between the work of~\citeasnoun{nob:fearn07} and our work is that while~\citeasnoun{nob:fearn07} work in the usual mixture setting where the observed data directly follows a finite mixture, in the LPCM the (latent) mixture data is actually something to be inferred. Clearly, this is quite different, since the latent mixture data is related to the observed data only through the logistic regression model. For convenience, the marginal posterior will be termed ``collapsed'' and the associated MCMC sampler in the next section, the collapsed sampler.

\subsection{ Estimation using MCMC }

Approximate sampling from the posterior (\ref{eq:post}) can be carried out using MCMC methods. There are four types of updates in our collapsed sampler
\begin{enumerate}[(i)]
\item updating the abundance parameter $\beta$ from the observed data likelihood
\item updating the latent positions of actors $\x_1,\dots,\x_n$
\item updating actor labels $c_1,\dots,c_n$ in the finite mixture prior
\item updating the number of components $G$ in the mixture, by absorbing components or  ejecting new ones. 
\end{enumerate}

\subsubsection{ Update for $\beta$} \label{sec:betaupdate}
The intercept parameter is updated using a random walk Metropolis-Hastings step. A proposal value $\beta^*$ is drawn from a $\mathcal{N}(\beta,\sigma_{\beta}^2)$  distribution, where $\beta$ is the current value of the intercept in the chain. The proposed value is accepted with probability \[\min\left[1, \frac{p(\Y|\bX, \beta^*)\,\pi(\beta^*)}{p(\Y|\bX, \beta\,)\,\pi(\beta\,)}\right].\]
Note that  the calculation of $p(\Y|\bX, \beta)$ is an $O(n^2)$ computation. This is a major drawback when considering the potential applicability of the LPCM in larger networks. 
Some approaches have been proposed in the literature to circumvent this bottleneck, most notably, the case-control approximation of~\citeasnoun{rafteryetal2012}. We do not consider this problem explicitly in this paper, however, we do note that the log of the likelihood (\ref{eq:like}) is
\begin{equation}
\log p(\bY|\bX, \beta) = \sum_{(i,j) \in \mathcal{D}} \log \Pr\left( Y_{ij} = y_{ij} | \x_i, \x_j, \beta \right). \label{eq:loglike}
\end{equation}
The calculation of this sum (\ref{eq:loglike}) is {\it embarrassingly parallelizable} i.e. the sum over pairs $(i,j) \in \mathcal{D}$ may be split over $P$ available processors at the time of compute giving in good cases a factor $P$ reduction in compute times for the $\beta$ update. This could be a suggested approach to assuage the quadratic order calculation. Of course, the practicalities of parallelization mean that a favourable increase in efficiency will be implementation and example dependent.

\subsubsection{ Update for latent positions}

The latent positions are updated once each per sweep of the MCMC algorithm using a random walk Metropolis-Hastings update. For actor $i, i=1,\dots,n$, a new $\x_i^*$ is proposed from a $\mathcal{N}(\x_i, \sigma_{\x}^2 \I )$ distribution, where $\x_i$ is the current position of actor $i$ in the latent space. The updated value is accepted with probability
\[
\min\left[ 1, \frac{p_i(\Y|\bX^*,\beta)\,\lambda_{c_i}( \bX^*, \bc) }{p_i(\Y|\bX\,,\beta)\,\lambda_{c_i}( \bX\,, \bc)} \right]
\] 
where
\[
p_i(\Y|\bX,\beta) = \prod_{j \ne i} \Pr( Y_{ij} = y_{ij} | \x_i, \x_j, \beta )
\]
if the network is undirected and
\[
p_i(\Y|\bX,\beta) = \prod_{j \ne i} \Pr( Y_{ij} = y_{ij} | \x_i, \x_j, \beta ) \,\Pr( Y_{ji} = y_{ji} | \x_i, \x_j, \beta )
\]
if directed.

\subsubsection{ Updates for actor labels }

The label of each actor is sampled from its full conditional
$
\pi(c_i|\bc_{-i},\bX, G)
$ 
in a Gibbs step in each sweep of the algorithm. There is the possibility of label switching due to the non-identifiability of the mixture prior. This will be discussed further in Section~\ref{sec:invariance}. These Gibbs moves may only move one actor at a time between components. Moves which can move many actors at a time between clusters are also used. These follow the general prescriptions of \possessivecite{nob:fearn07} moves M1, M2 and M3.  As demonstrated by~\citeasnoun{nob:fearn07} (Section 3.4), such moves can improve the mixing of the chain. 

\subsubsection{Updating the number of components in the mixture prior}

The moves to update the number of components in the mixture comprises two reversible {\it eject} and {\it absorb} moves. If the current number of clusters is $G$, then it is proposed to eject a component from one of the existing components with probability $\eta_G^{\mathrm{ej}}$; the probability of proposing an absorb move is $1-\eta_G^{\mathrm{ej}}$. For all $G$ except 1 and some maximum realistic number $G_{\mathrm{max}}$ components, we use $\eta_G^{\mathrm{ej}} = 0.5$.

The eject move chooses one of the $G$ existing clusters $g$ at random. It will be attempted to potentially reallocate members of $g$ to a new component $G+1$. A probability $p$ is sampled from a beta $\mathcal{B}(a,a)$ distribution. The elements of component $g$ are each put into component $G+1$ with probability $p$. The value of $a$ is chosen from a precomputed lookup table, so that ``empty components are proposed relatively often'' (see \citeasnoun{nob:fearn07}, \citeasnoun{wyse:friel12}). The proposal mechanism creates a new label vector $\bc^* \in \{1,\dots,G+1\}^n$ resulting in the acceptance probability $\min[1,\rho]$, where 
\[
\rho = \frac{\lambda_{g}(\bX,\bc^*) \lambda_{G+1}(\bX,\bc^*) \pi(G+1)}{\lambda_g(\bX,\bc) \pi(G) }\,\,\frac{1-\eta_{G}^{\mathrm{ej}}}{\eta_{G}^{\mathrm{ej}}} \,\,\frac{\Gamma(a)^2}{\Gamma(2a)} \frac{ \Gamma( 2a + n_g)}{\Gamma(a + n_g^*)\Gamma(a + n_{G+1}^*)}.
\]
If the move is accepted a random label swap is made between component $G+1$ and one of the other components. 

In proposing an absorb move, two components $g$ and $k$ are selected at random from the $G+1$ available. Suppose that the current label vector is $\bc$. It is proposed to combine these into one component, in other words, $g$ absorbs $k$ if $g<k$ and vice-versa. Actors which are labelled $k$ are relabelled $g$, giving the proposed label vector $\bc^*$. Then the move is accepted with probability $\min[1,\upsilon]$ where
\[
\upsilon = \frac{ \lambda_g(\bX, \bc^*) \pi(G)}{\lambda_g(\bX,\bc)\lambda_k(\bX,\bc)\pi(G+1)} \,\,\frac{\eta_{G+1}^{\mathrm{ej}} }{ 1-\eta_{G+1}^{\mathrm{ej}} }\,\, \frac{\Gamma(2a)}{\Gamma(a)^2} \frac{\Gamma(a+n_g)\Gamma(a+n_k)}{\Gamma(2a + n_g^* )},
\]
and $n_g^* = n_g + n_k$. If the move is accepted, all elements of the label vector with a value of $k$ upwards are decremented by 1. 

\subsection{Model invariance and post-processing} \label{sec:invariance}

By close inspection, it can be seen that the likelihood given by (\ref{eq:like}) is invariant to rotations, reflections or translations of the latent positions $\bX$. This is because the linear predictor (\ref{eq:linpred}) depends only on the distance between the latent positions.  When computing estimates of posterior quantities involving the latent positions via ergodic averages it is thus necessary to post-process the samples generated by the MCMC algorithm. A Procrustes transformation \cite{sibson1978} is used to match each sample to a reference set of positions $\bX_{\mathrm{ref}}$. The MCMC sample iterate giving the highest likelihood (\ref{eq:like}) is used as a reference configuration.

Additionally it can be seen that another kind of invariance is present in the mixture prior (\ref{eq:mixprior}). Any permutation $\sigma$ of $\{1,\dots,G\}$  applied to the labels $\bc$ will produce the same value of the prior i.e. $\pi(\bX, \bc|\bftheta) = \pi(\bX,\bc_{\sigma}|\bftheta)$ where $\bc_\sigma = (\sigma(c_1),\sigma(c_2),\dots,\sigma(c_n))$. Again, to estimate posterior functionals of the labels, the samples of labels must be post-processed. To do this we use an iterative square assignment algorithm which is detailed in full in Appendix C of~\citeasnoun{wyse:friel12}. This algorithm finds the best permutation for each sample by minimizing a cost function based on component assignment agreement.

\subsection{Incorporating uncertainty in hyperparameters} \label{sec:uncertainhp}

Exploration of the range of possible values of the hyperparameters can be important for some applications. Of the hyperparameters in the model, in our experience, $\gamma$ appears to be the one whose prior specification has the strongest influence on the posterior. This mirrors closely the findings of~\citeasnoun{richardson:green97} (Section 5.1), although their prior specification is slightly different to the one adopted here.  The posterior of the number of groups and the prior choice of $\gamma$ are closely connected, since $\gamma$ effectively controls the volume  in latent space that clusters can occupy. Small values place higher prior mass on clusters occupying a smaller volume of latent space (hence a higher number of groups), while large values favour a smaller number of groups. However, universal calibration of $\gamma$ is not possible for all problems {\it a priori}. Incorporating a hyperprior on $\gamma$ can mitigate this calibration issue. In an extra sampling step, the component marginal precisions $\tau_g$ can be ``uncollapsed'' and sampled at each iteration (still leaving the $\bfmu_g$ collapsed). Assuming a $\mbox{Gamma}(s/2,r/2)$ hyperprior for $\gamma$, first sample $\tau_g$ from the conditional
\[
\tau_g\, |\, G, \bc, \bX, \gamma \sim \mathcal{G}\left( \frac{n_g d + \delta}{2} \, , \, \frac{1}{2}\left[\sum_{i:c_i=g} || \x_i ||^2 - \frac{|| \sum_{i:c_i = g} \x_i ||^2}{n_g + \kappa}  + \gamma \right] \right)
\]
and then sample
\[
\gamma  \, | \, G, \tau_{1:G} \sim \mathcal{G}\left( \frac{G\, \delta + s}{2}\,,\,\frac{1}{2}\left[\sum_{g=1}^G \tau_g + r\right] \right)
\]
in each sweep of the algorithm. Uncertainty in $\kappa$ could also potentially be incorporated using this type of approach, whereby one would additionally sample the $\bfmu_g$ from their full conditionals in order to sample $\kappa$ (having assumed a hyperprior for it).

\section{Application to simulated networks} \label{sec:simulated}

\begin{figure}
\begin{center}
\includegraphics[width=50mm]{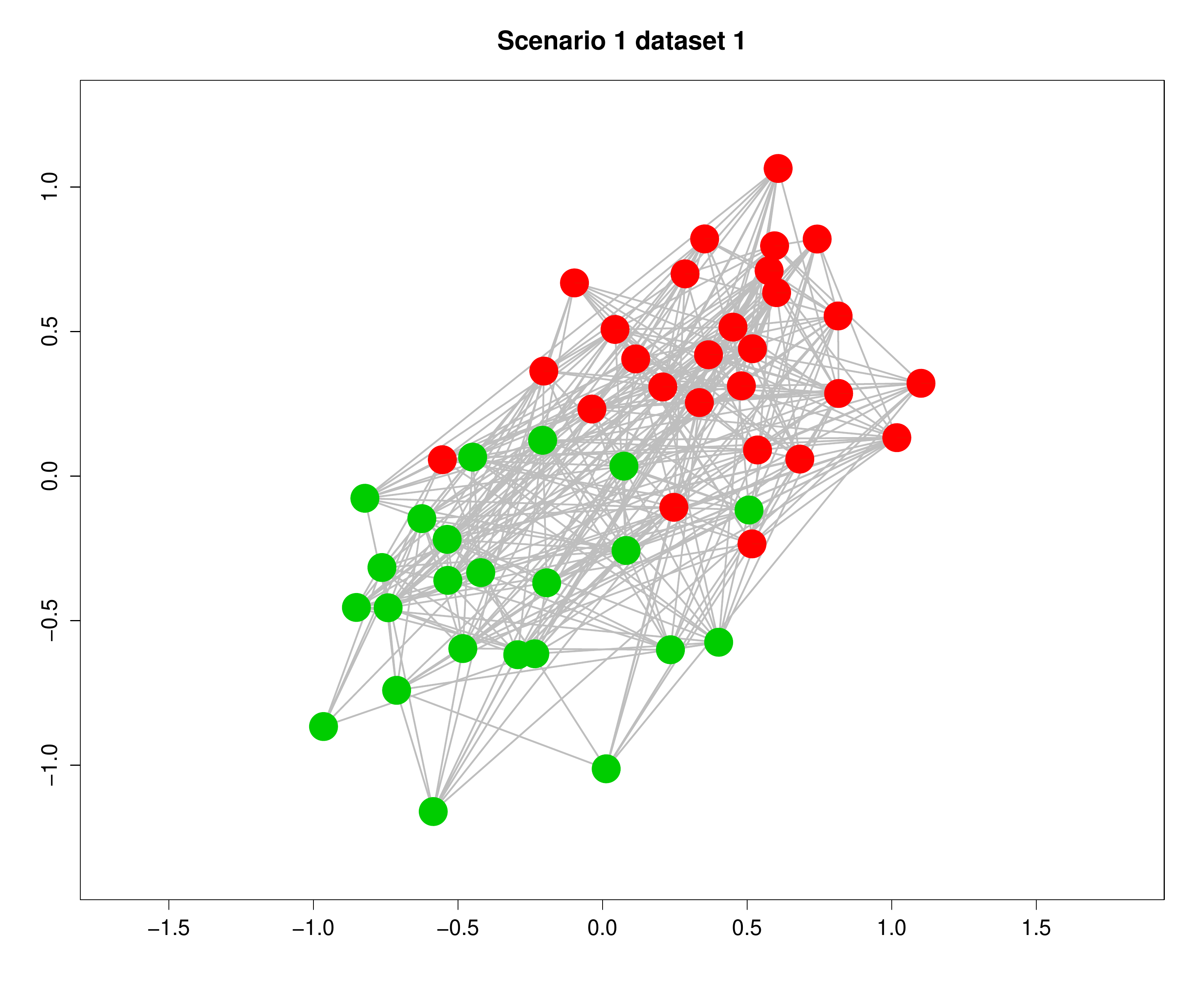}
\includegraphics[width=50mm]{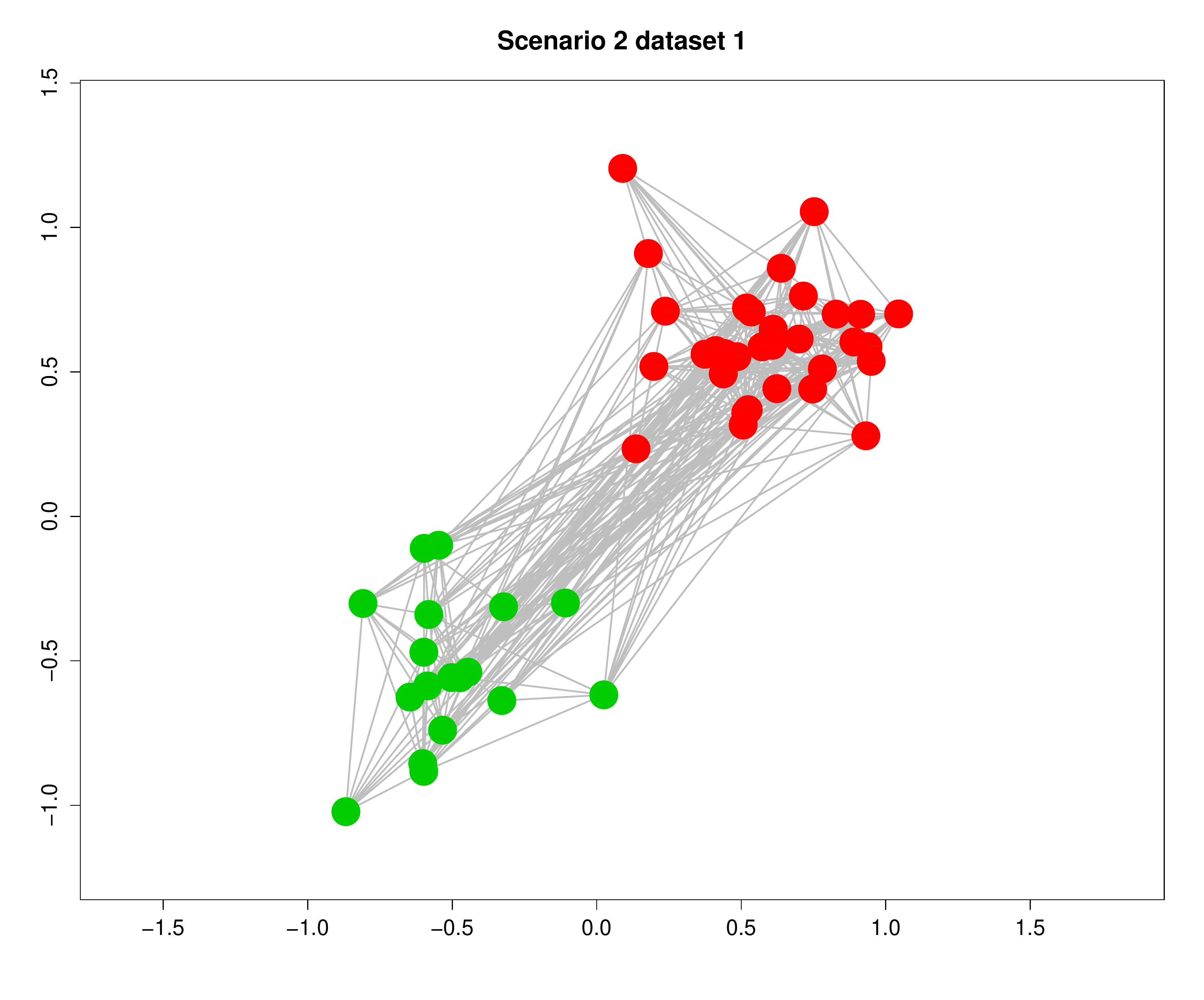}
\includegraphics[width=50mm]{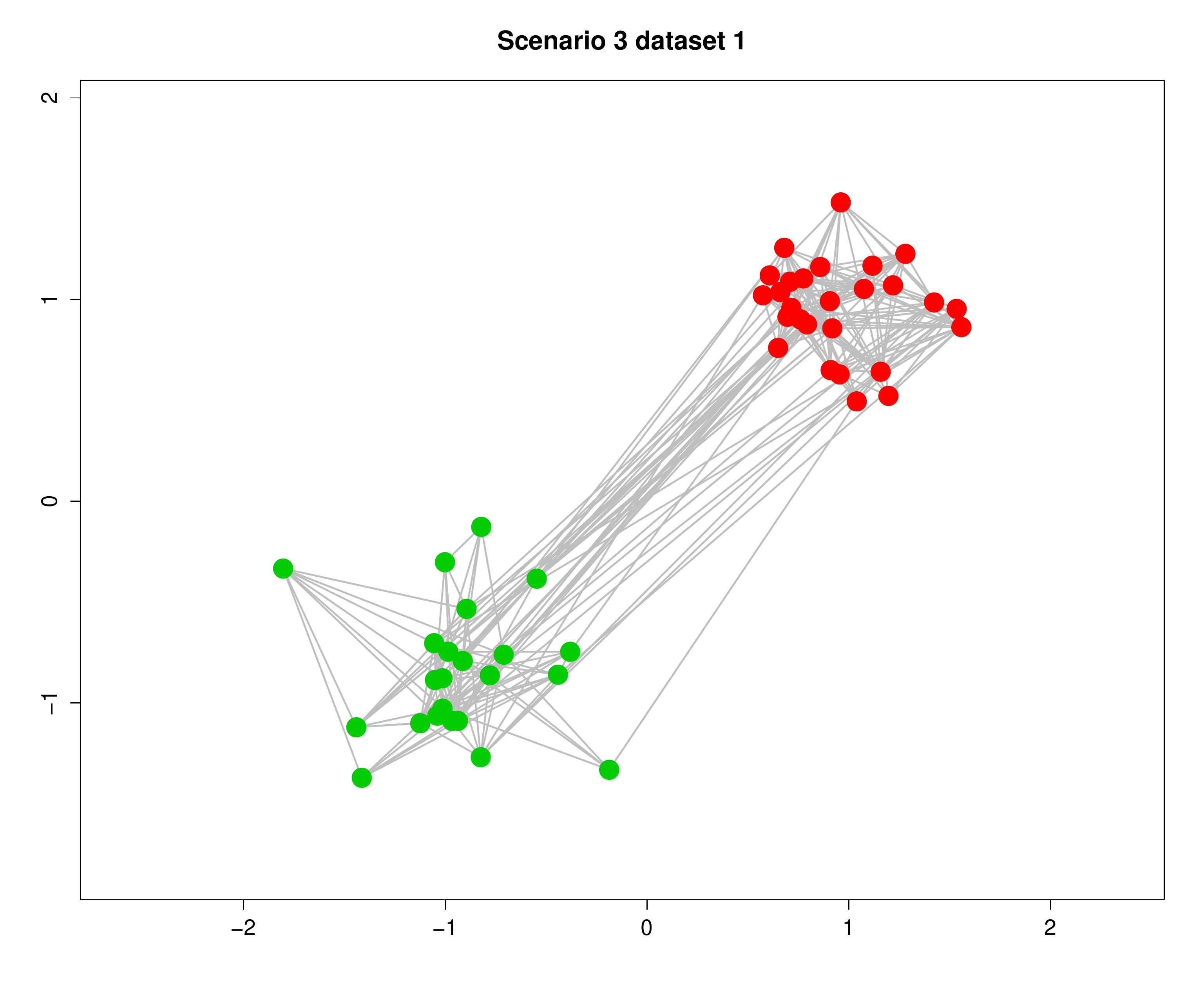}
\includegraphics[width=50mm]{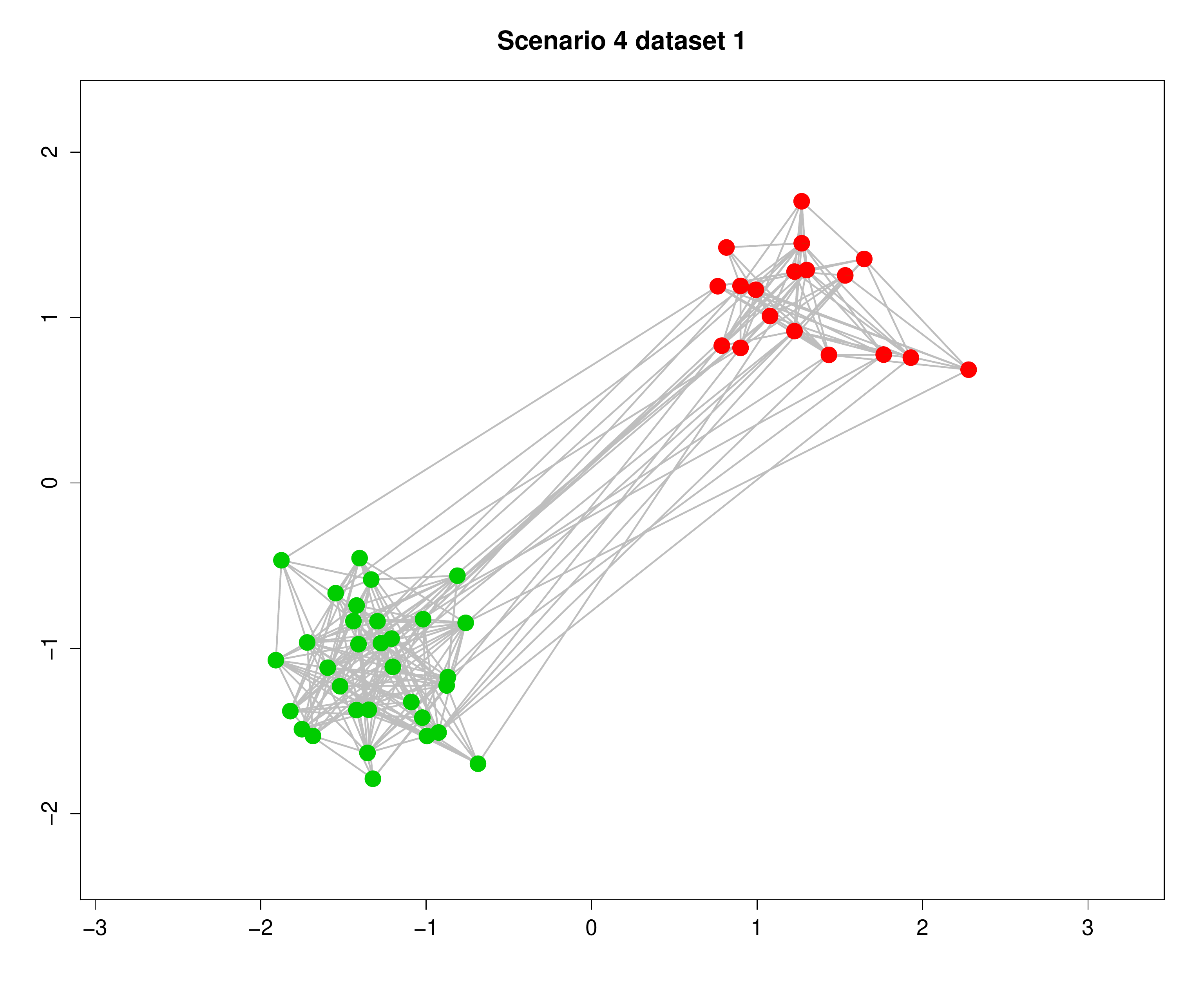}
\includegraphics[width=50mm]{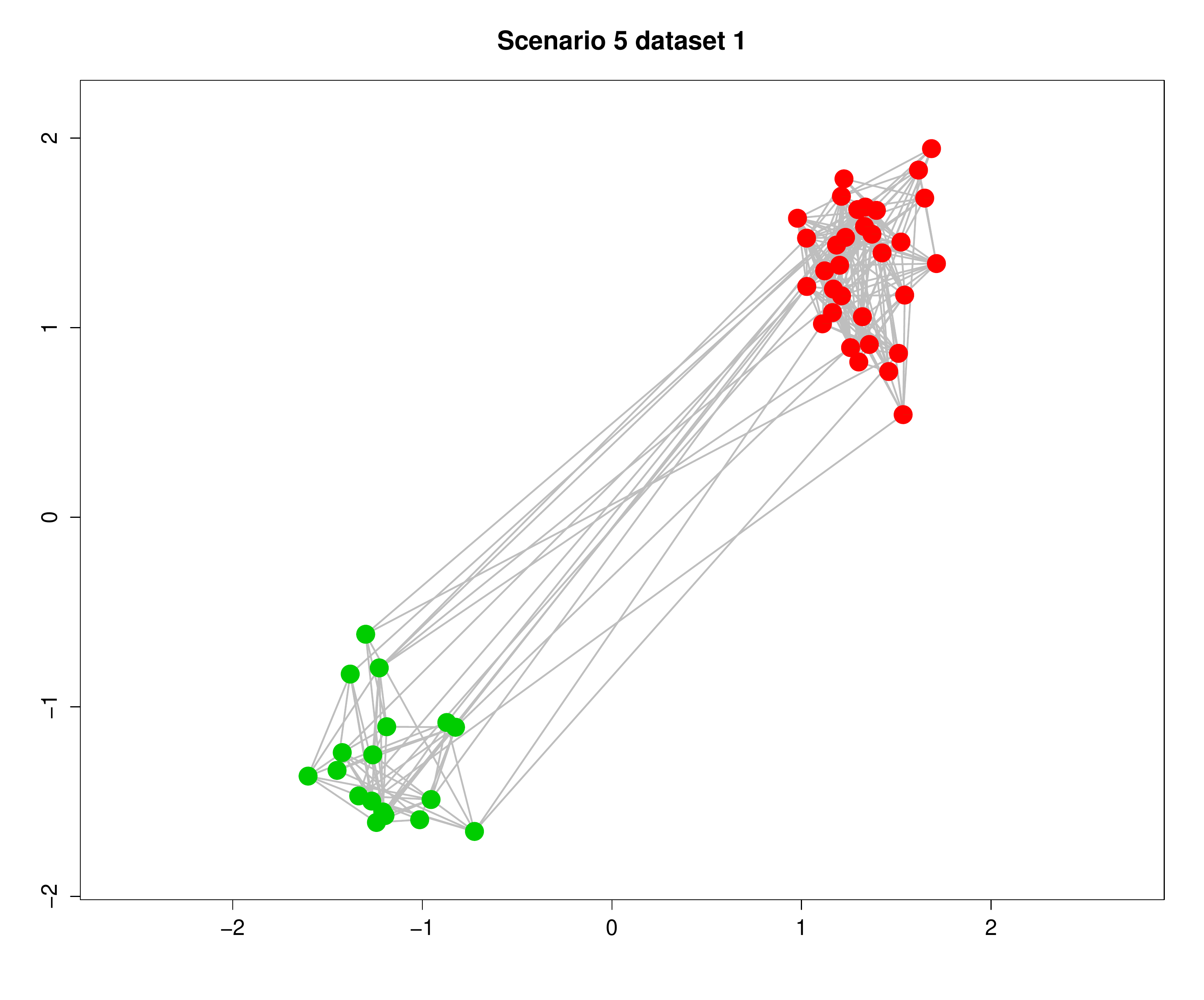}
\includegraphics[width=50mm]{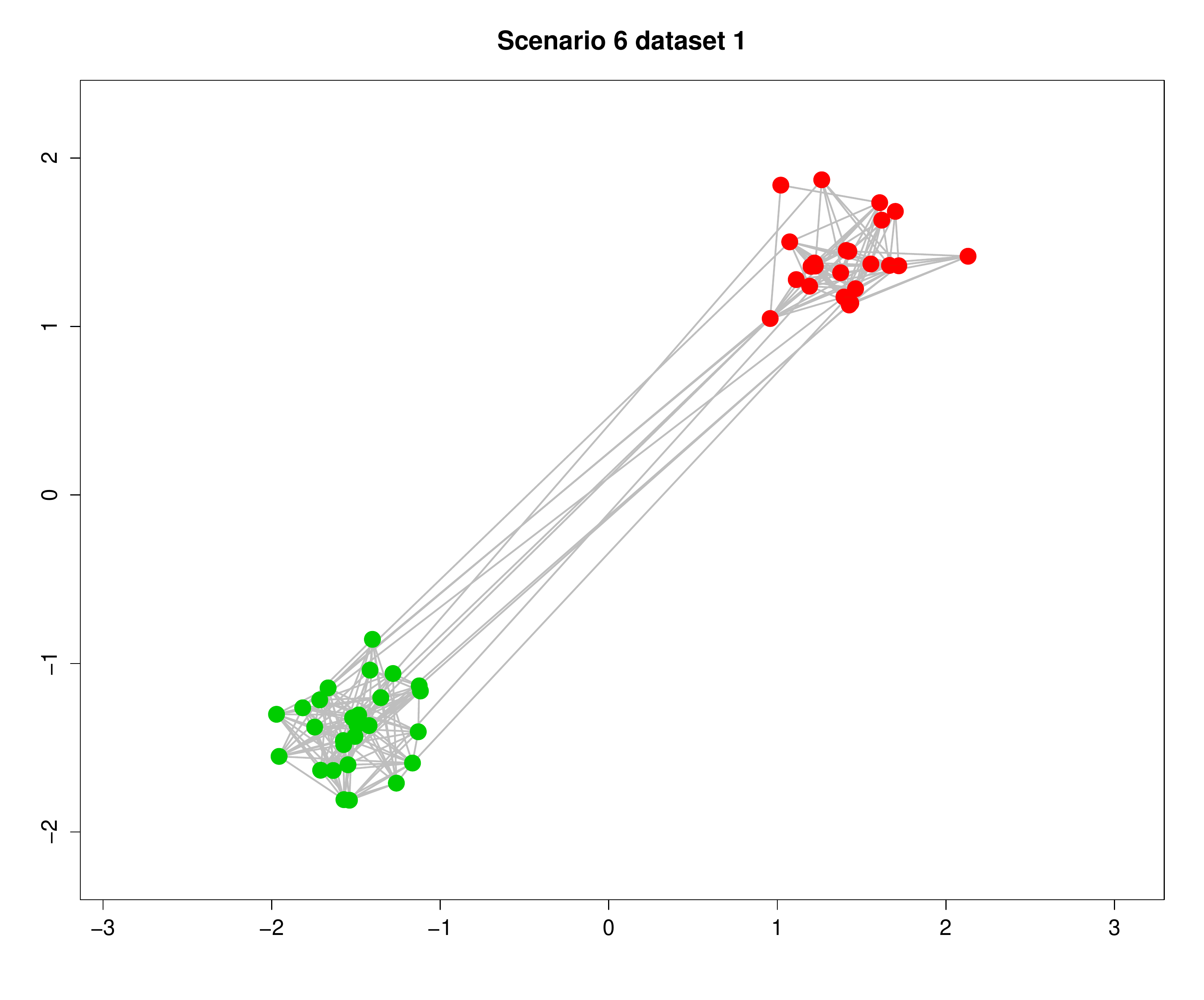}
\end{center}
\caption{Latent positions of simulated networks with links indicated in grey for Scenarios 1 to 6 respectively.} \label{fig:simlpos}
\end{figure}

Here we present a simulation study to benchmark our approach. We simulated 50 actor networks from a model with $G=2$ groups and cluster centres given by 
\[
\bfmu_1 = \bfmu = (\mu,\mu)^T, \qquad \bfmu_2 = -\bfmu =  (-\mu,-\mu)^T,
\]
cluster precisions $\tau_1 = \tau_2 = \tau$ and equal weights $\omega_1 = \omega_2 = 0.5$. Values of $\mu$ and $\tau$ were chosen to give different levels of separation and inter group connectivity indicated by parameter $r$ as described in Appendix~\ref{sec:appendix}, with $r = 1$ implying poor separation and larger values of $r$ implying more well separated clusters. Figure~\ref{fig:simlpos} shows the simulated latent space positions of six example simulated networks for the different scenarios $r \in \{1.5, 2.5,5,10,15,20\}$ (see Appendix~\ref{sec:appendix}) which we term scenarios 1-6 respectively. For each value of $r$ we simulated 100 networks in total and fitted the LPCM to approximate the posterior distribution of $G$ using a run of our MCMC algorithm in each case. Each run consisted of 10,000 burn-in iterations and a further 50,000 iterations, retaining every $10$th. We used hyperparameter values as described in Section~\ref{sec:examples}.

\begin{figure}[!h]
\begin{center}
\includegraphics[width=50mm]{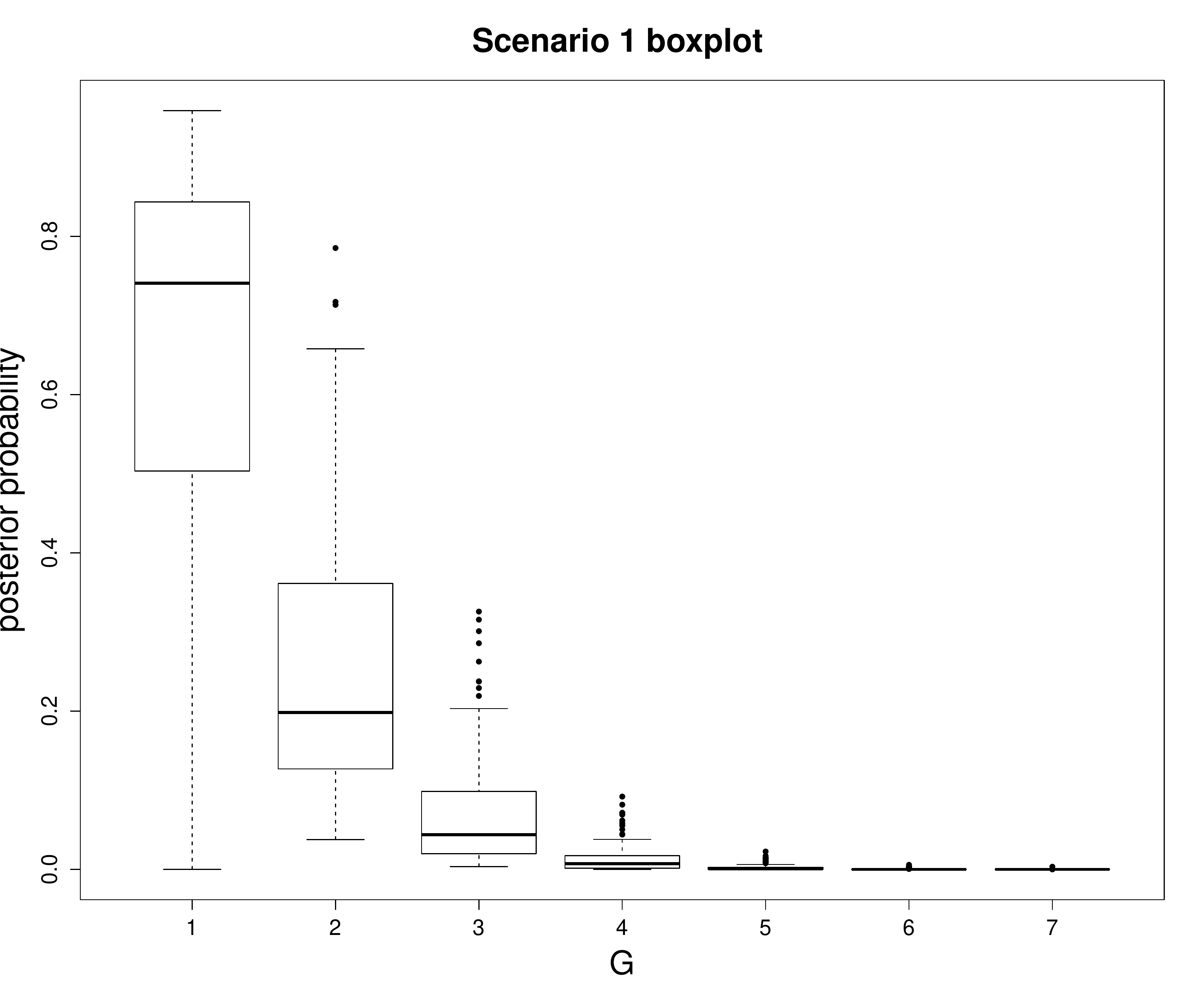}
\includegraphics[width=50mm]{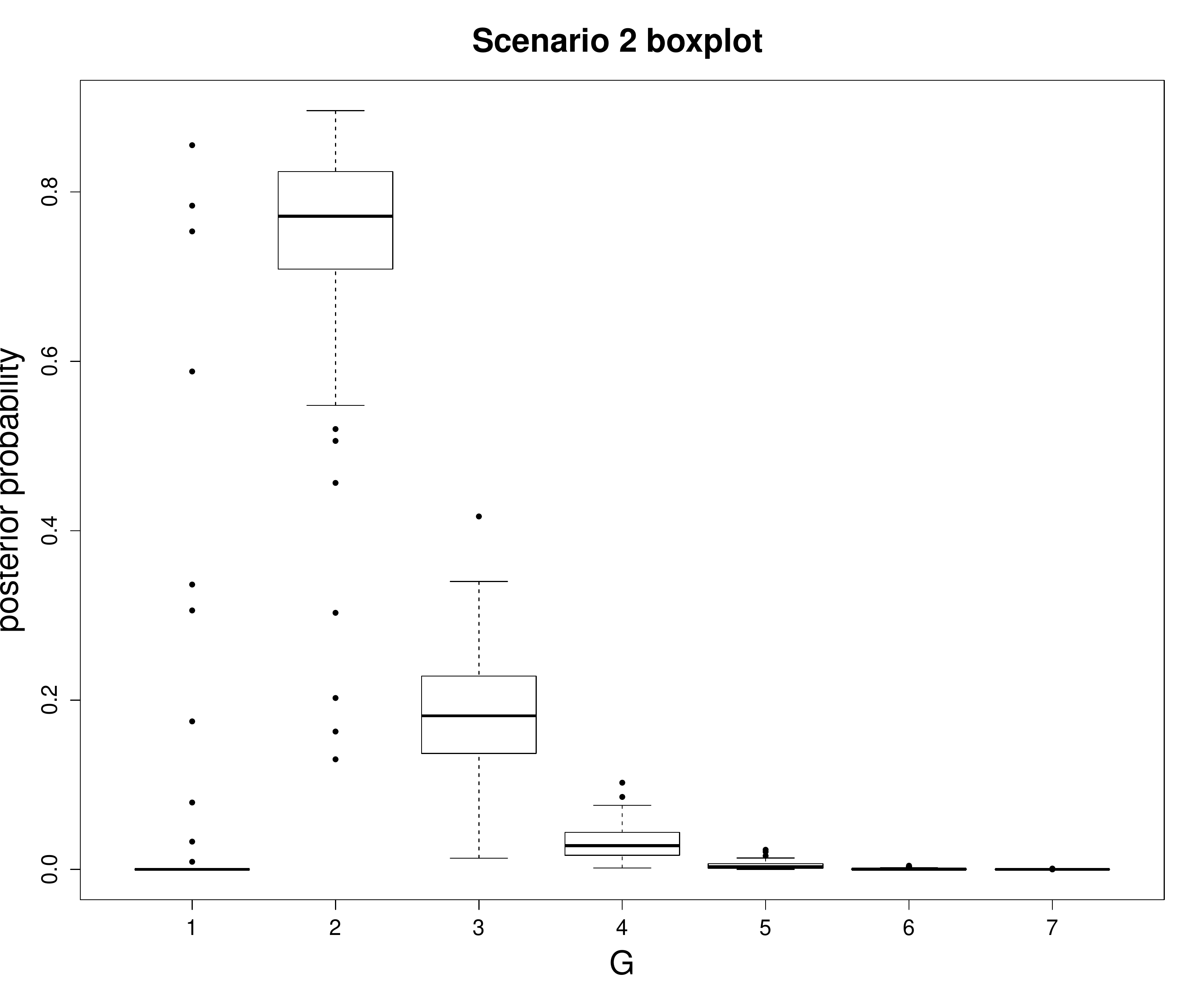}
\includegraphics[width=50mm]{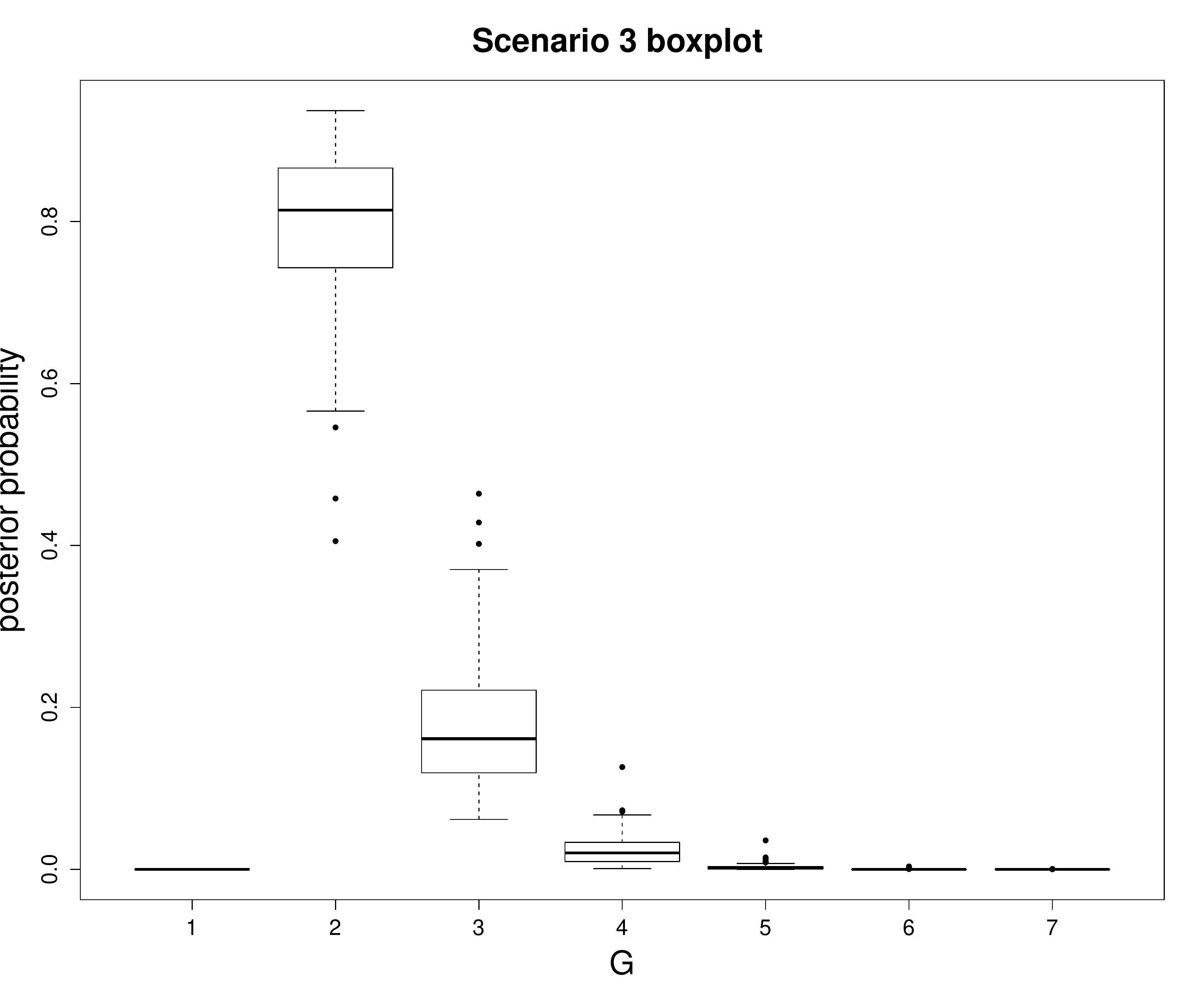}
\includegraphics[width=50mm]{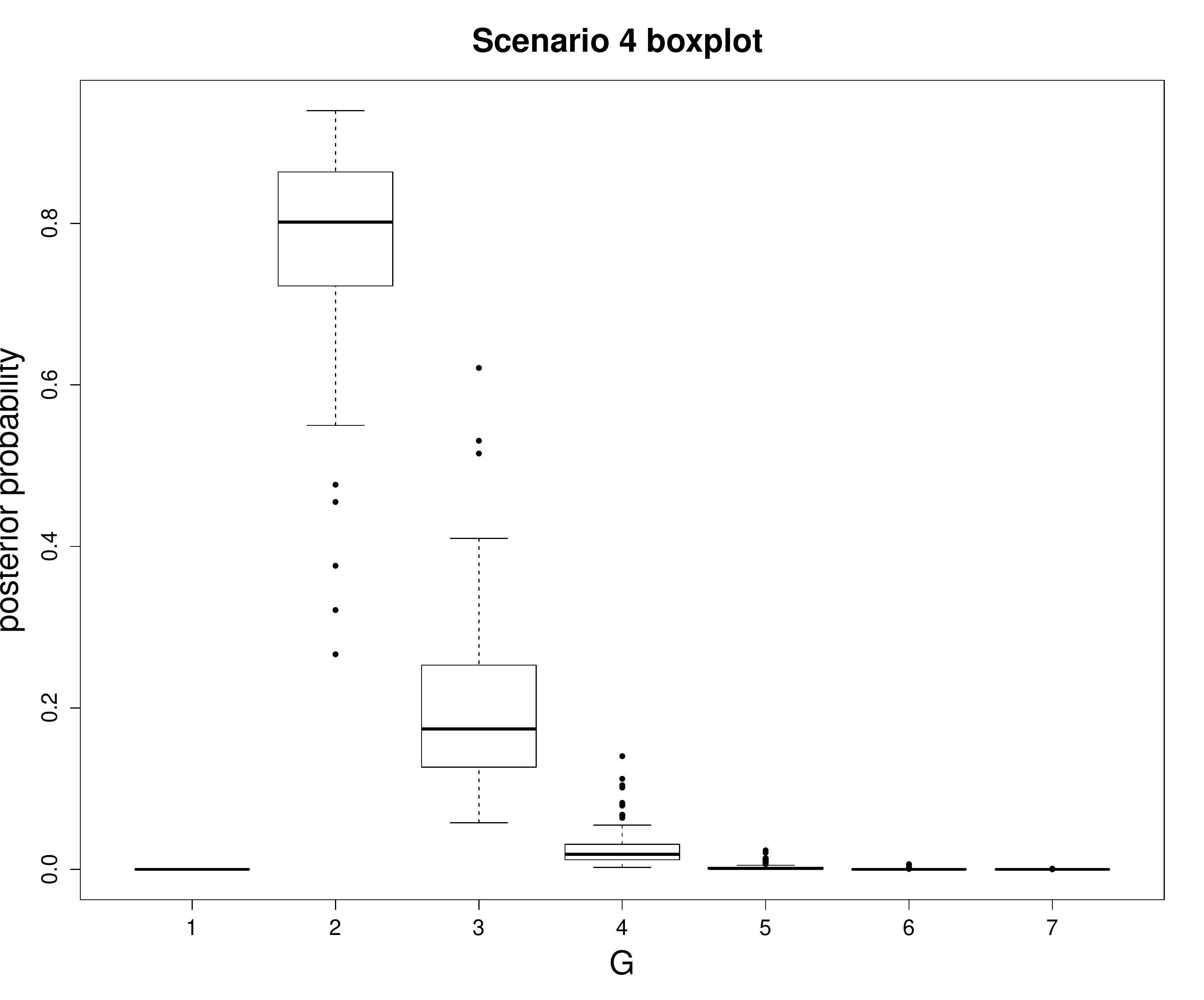}
\includegraphics[width=50mm]{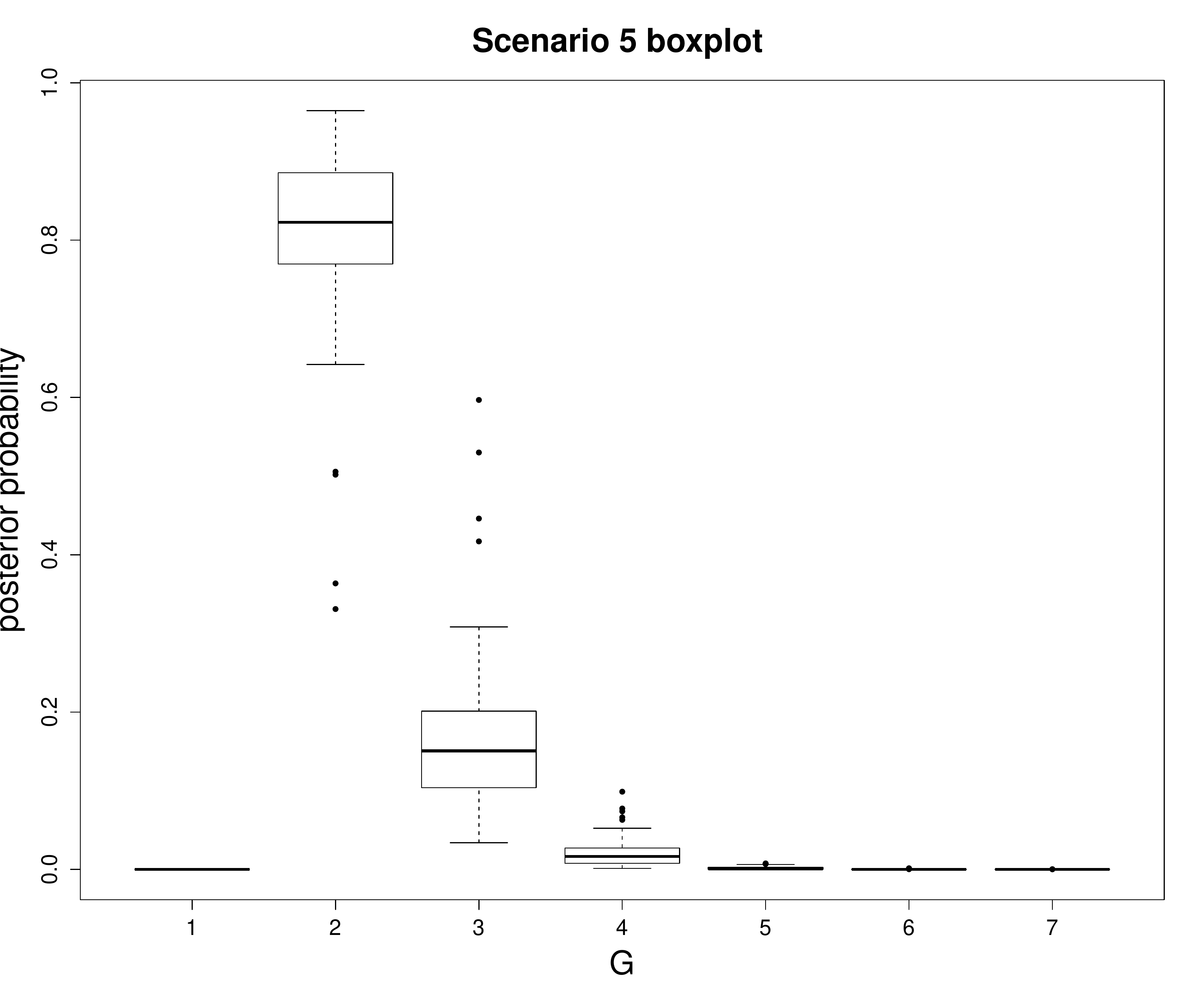}
\includegraphics[width=50mm]{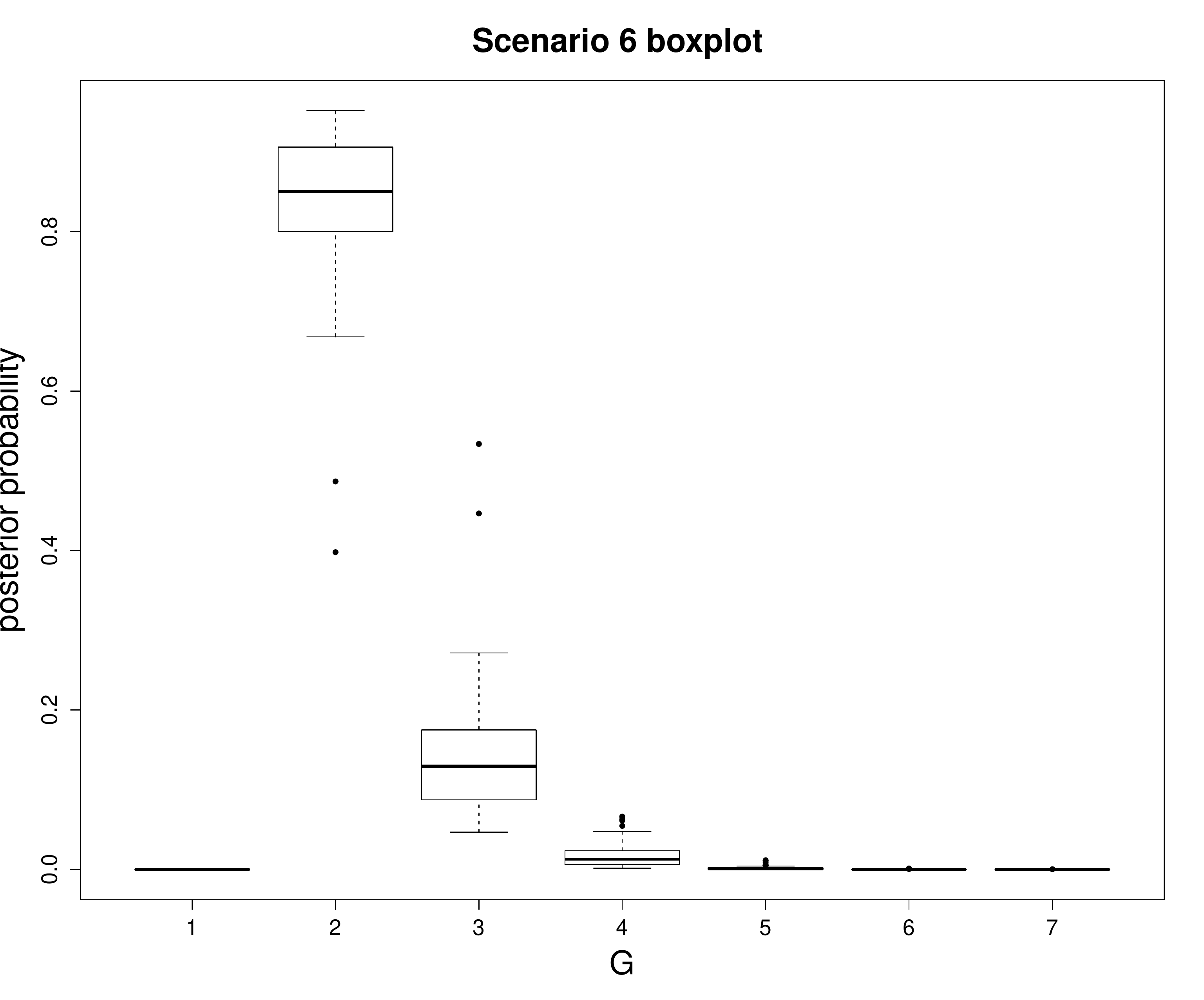}
\end{center}
\caption{Boxplots showing posterior distribution of $G$ over 100 siumulated networks for Scenarios 1 to 6 respectively.} \label{fig:simboxplot}
\end{figure}

We expect that as $r$ increases, the true value of $G$ becomes easier to identify. This is what we see in Figure~\ref{fig:simboxplot}, which shows boxplots of the estimated posterior probability of $G$ from our sampler over the 100 simulated networks for each scenario. For $r=1.5$ where the clusters are close, $G=1$ identified with high probability in most cases. For $r>2.5$, $G=2$ is correctly identified with increasing (with $r$) probability.

\section{Application to real data} \label{sec:examples}

We now illustrate our approach using some well known social networks, Sampson's $18$ node network \cite{sampson68},  Zachary's $34$ node karate club network \cite{zachary77} and a $62$ node network of New Zealand Dolphins \cite{lusseauetal03}. The settings for our algorithm for all applications below are to take $\alpha = 3, \delta = 2$ and $\kappa = .1$. The value of $\gamma$ is sampled using the approach outlined in Section~\ref{sec:uncertainhp}. The values of hyperprior parameters $s$ and $r$ are chosen so that the prior mean of $\gamma$ is 0.103 with a standard deviation of $0.103/4$. The proposal standard deviations $\sigma_{\beta}$ and  $\sigma_{\x}$ are chosen to give an aggregated $25-40\%$ acceptance rate of proposed moves (aggregated over all latent positions). We refer to our sampler as the \texttt{collapsed} sampler.

The examples serve to illustrate model uncertainty for well known social networks and to make comparisons with inference using \texttt{latentnet} \cite{Kriv:Hand07,Kriv:Hand13} and the variational approximation to the posterior using \texttt{VBLPCM}~\cite{salter:murphy12}. Inference using \texttt{latentnet} involves sampling  from the full posterior of~\citeasnoun{handcocketal07}. The number of clusters $G$ is fixed and inference is carried out separately for $G=1,\ldots,G_{\mathrm{max}}$. The approximative BIC is used to choose the `best' value of $G$ which is most supported by the data. The Variational Bayes approach to inference is implemented using \texttt{VBLPCM} \cite{salter:murphy12}. 
Inference is carried out separately for $G=1,\ldots,G_{\mathrm{max}}$ component models. A good initialisation of the variational parameters is important \cite{salter:murphy12}. The Fruchterman-Reingold layout is used to initialize the latent positions, followed by the use of \texttt{mclust} \cite{fraley:raftery02,fraley:raftery03}
to initialize the clustering parameters. 
The Fruchterman-Reingold layout algorithm is itself initialized using a random configuration, thus introducing a stochasticity into the algorithm and different results will be found each time. The variational approximation which is `closest' to the true posterior in terms of Kullback-Leibler divergence is chosen as the best $G$ component model from $10$ different initialisations. The approximate BIC discussed in Section \ref{bicsec} is then used to choose the number of components. 

\subsection{Sampson's monks}\label{resmonks}

Sampson \citeyear{sampson68} conducted a social science study of  $18$ monks in a monastery during the time of Vatican II. During the study, a political `crisis in 
the cloister' resulted in the expulsion of four monks and the voluntary departure of several others.
We use the aggregated version of this network widely used in the social network analysis literature.

\begin{figure}
\begin{center}
\includegraphics[width=80mm]{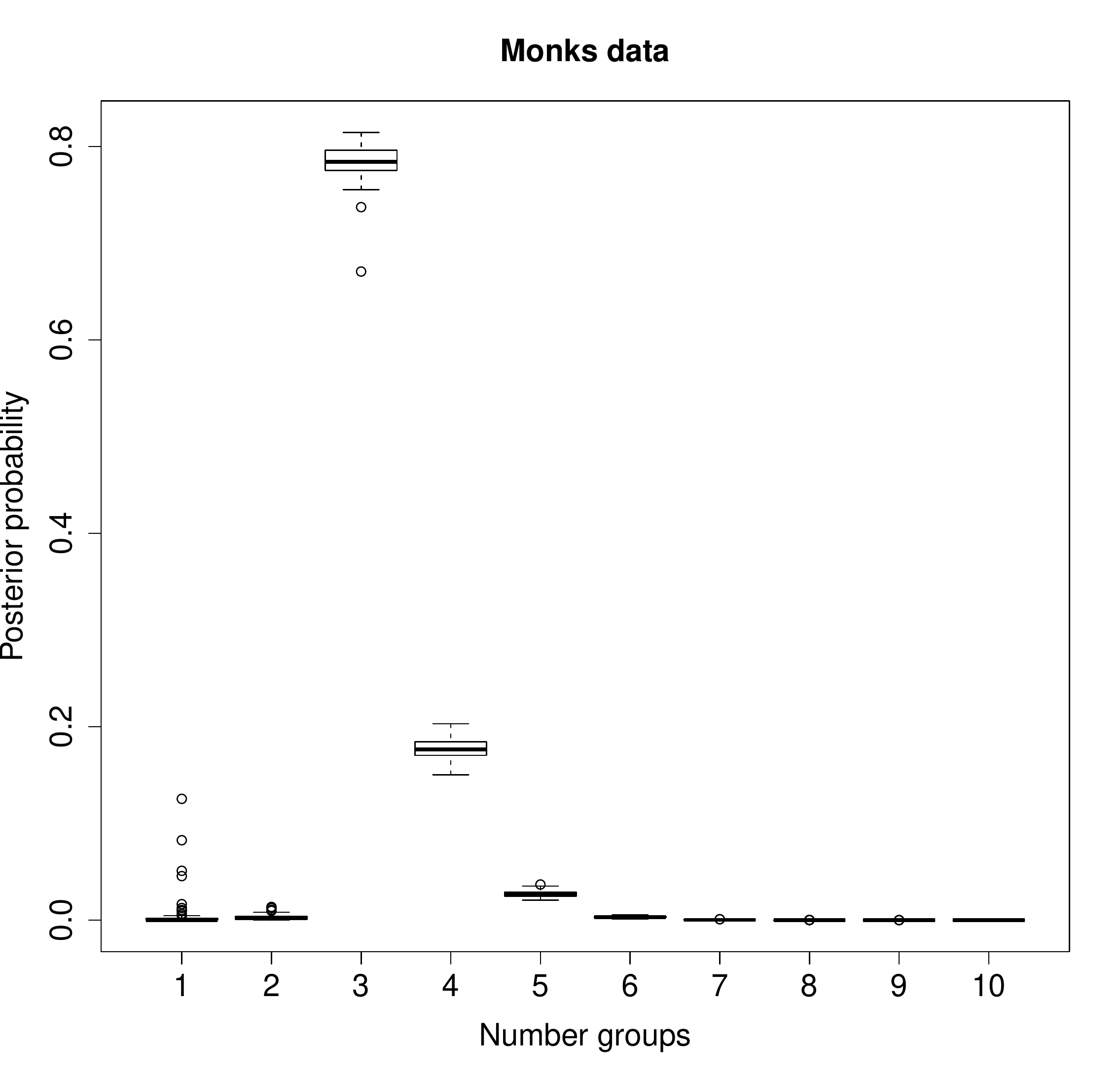}
\end{center}
\caption{Barplot showing uncertainty over $G$ for 100 runs of the MCMC sampler on the Monks data.} \label{fig:monkbarplot}
\end{figure}

\begin{figure}
\begin{center}
\includegraphics[width=150mm]{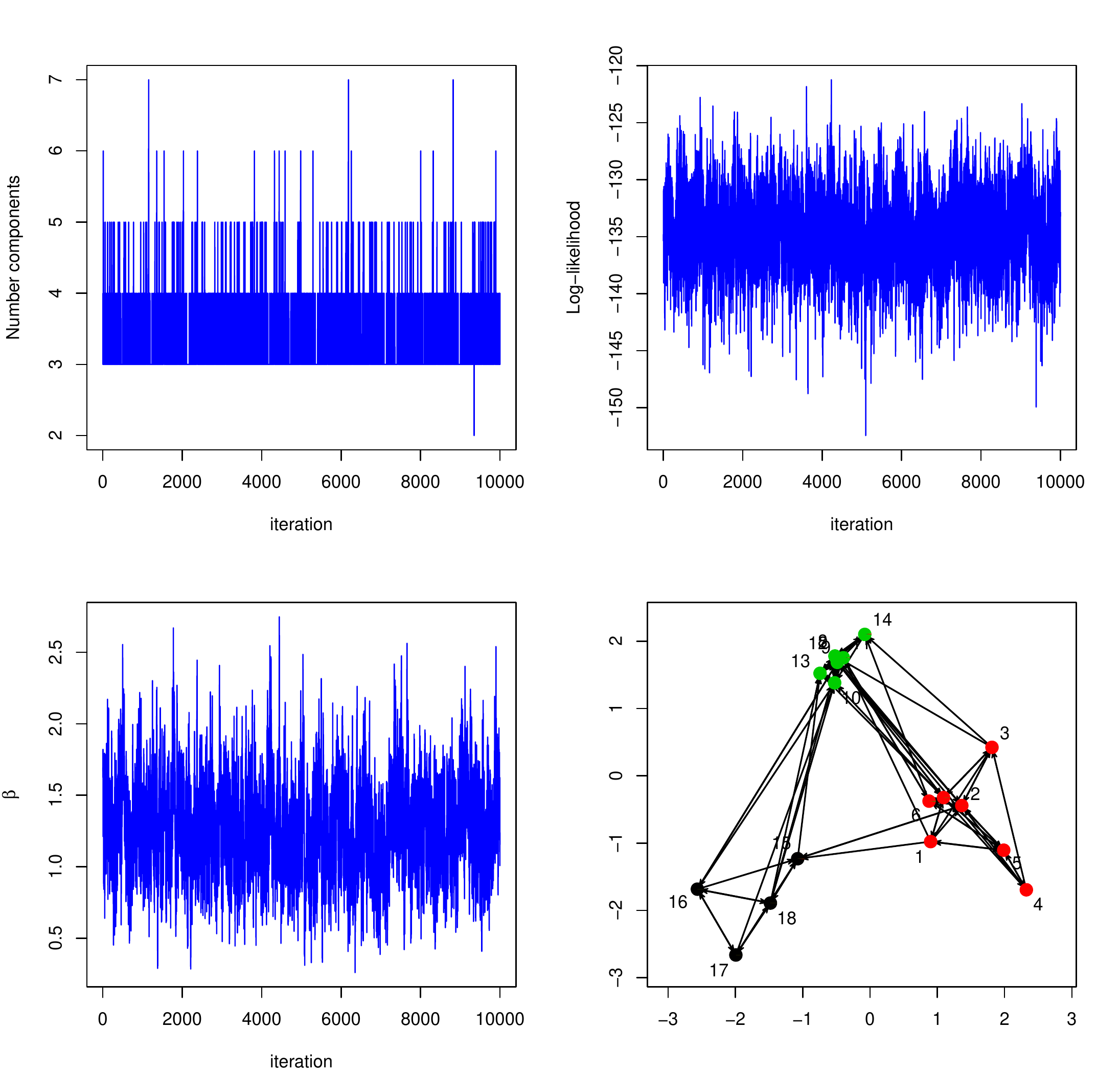}
\end{center}
\caption{Summary of MCMC sample for the Monks data. Left-right, top-bottom; traceplot of sampled number of components, log-likelihood, intercept. Bottom right, estimated posterior mean positions for a three group model, coloured by group label.}\label{fig:monksres}
\end{figure}

Figure~\ref{fig:monksres} shows a summary of a run of 100,000 iterations of the \texttt{collapsed} sampler (having discarded 10,000 burn-in), and retaining every 10th iterate. The eject/absorb moves had an acceptance rate of 4\%. The uncertainty in the number of groups in the monastery becomes clear from the top left trace plot for sampled values of $G$. A $3$ or $4$ component model is widely accepted as the most suitable clustering for this data. Figure~\ref{fig:monkbarplot} shows the uncertainty in the number of groups quantified by running the MCMC algorithm above 100 times and examining the distribution of posterior probabilities of given number of components, showing agreement with the general consensus on the number of groups. 

The results of our analysis for Sampson's monks network are qualitatively quite similar to the inference using \texttt{latentnet} shown in~\citeasnoun{handcocketal07} (Section 5.1).
As described in Section \ref{lpcm}, differences in prior specification should be kept in mind when 
comparing inference using the collapsed sampling, inference using \texttt{latentnet} and the variational approach using \texttt{VBLCPM}.

\begin{table}[htp]
\begin{center}
\begin{tabular}{| l|c|c|c|c|c|}
\hline
 & $G=1$ & $G=2$ & $G=3$ & $G=4$ & $G=5$ \\
\hline
	\texttt{collapsed} & 0.01 & 0.01 & \bf{0.77} & 0.18 & 0.03 \\
  \texttt{latentnet} BIC & 380.87 & 373.36 & \bf{336.49} & 342.12 & 347.66 \\ 
  \texttt{VBLPCM} BIC & 540.98  & 504.67   & \bf{477.62}  & 490.78 & 514.67\\ 
  \hline
\end{tabular}
\caption{Estimated posterior distribution of $G$ from our \texttt{collapsed} sampler, and approximate BIC using \texttt{latentnet} and \texttt{VBLPCM} for the Monks network. Favoured model is indicated in bold font.}
\label{tab:pallmonks}
\end{center}
\end{table}

Qualitatively different results were seen for the variational approximation using \texttt{VBLPCM} 
with less separation of clusters and practically no uncertainty in cluster membership. One drawback of the variational approach is that the divergence 
between the two distributions can only be quantified up to an unknown constant of proportionality, which could affect the approximation quality significantly.

\begin{figure}
 \centering
\includegraphics[width=7cm]{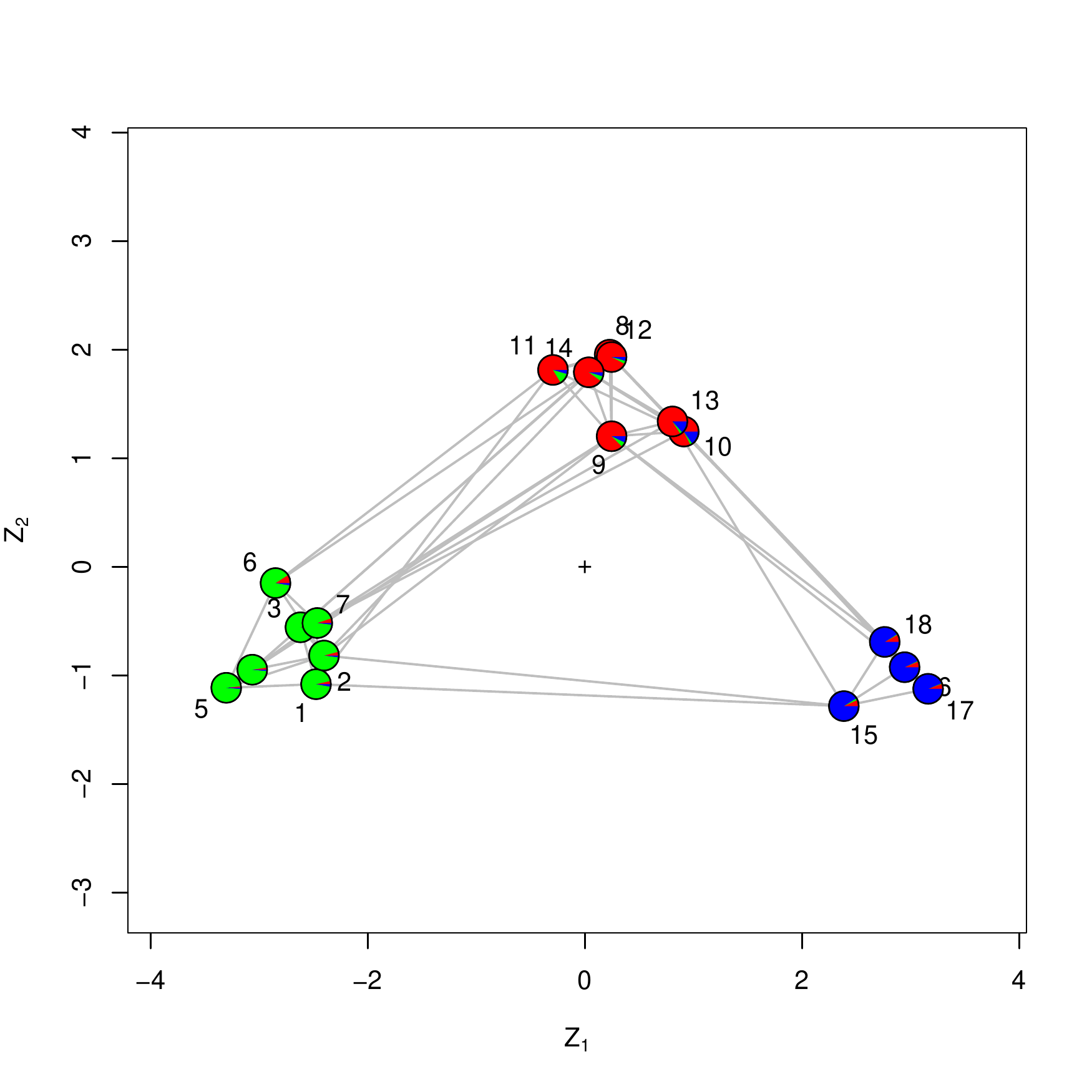}
\includegraphics[width=7cm]{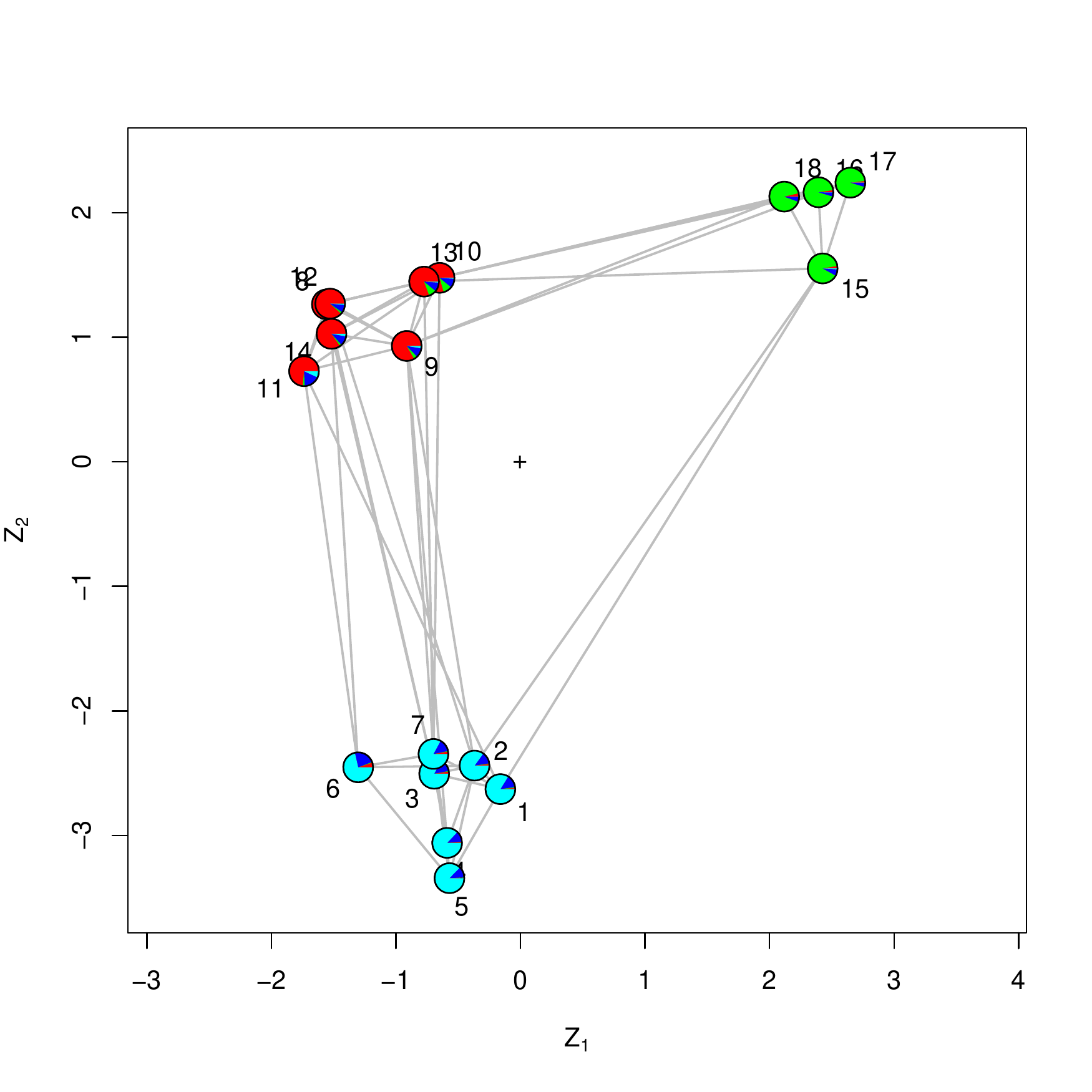}
\caption{Posterior mean latent positions for a $3$ and $4$ component model for Sampson's monks using \texttt{latentnet}.}
\label{lnetmonks}
\end{figure}

\pagebreak

\subsection{Zachary's Karate Club}\label{sec:zachary}

Zachary's karate club  \cite{zachary77} consists of $78$ undirected friendship ties between $34$ members of  a karate club. 
The club split due to a disagreement between the club president and the coach, both of whom are included in the network as actors $1$ and $34$ respectively. 

\begin{figure}
\begin{center}
\includegraphics[width=80mm]{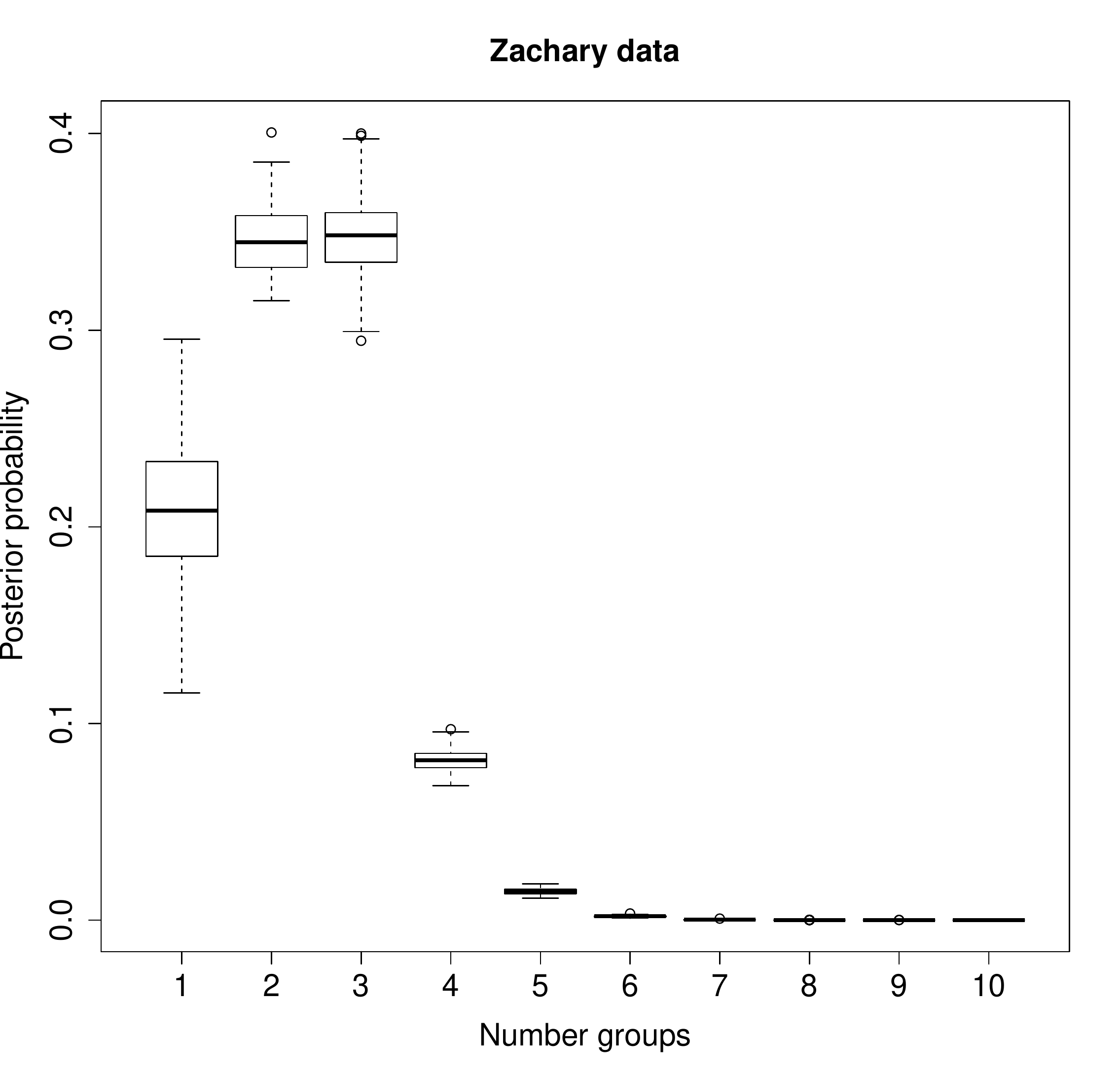}
\end{center}
\caption{Barplot showing uncertainty over $G$ for 100 runs of the MCMC sampler on the Zachary network.} \label{fig:zacbarplot}
\end{figure}

The coach formed a new club with some of the members.
It is interesting to compare the actual split of the club and the clustering of the friendship network.
This is another example of a dynamic network which is usually examined in a static aggregated form in the social network analysis literature. 

Figure~\ref{fig:zacbarplot} shows the results of 100 runs of the sampler, showing similar support for both a two and three group model. Notably, there is also appreciable posterior support for no clustering in the network. 
Approximate BIC values given by \texttt{latentnet} and \texttt{VBLPCM} are displayed in Table \ref{tab:pallzac} along with posterior probabilities from our sampler.  
The latent positions and the intercept mixed well
using proposal variances $\sigma^2_{\x}=1.7$ and $\sigma^2_\beta=0.5$. 

\begin{figure}
 \begin{center}
\includegraphics[width=7cm]{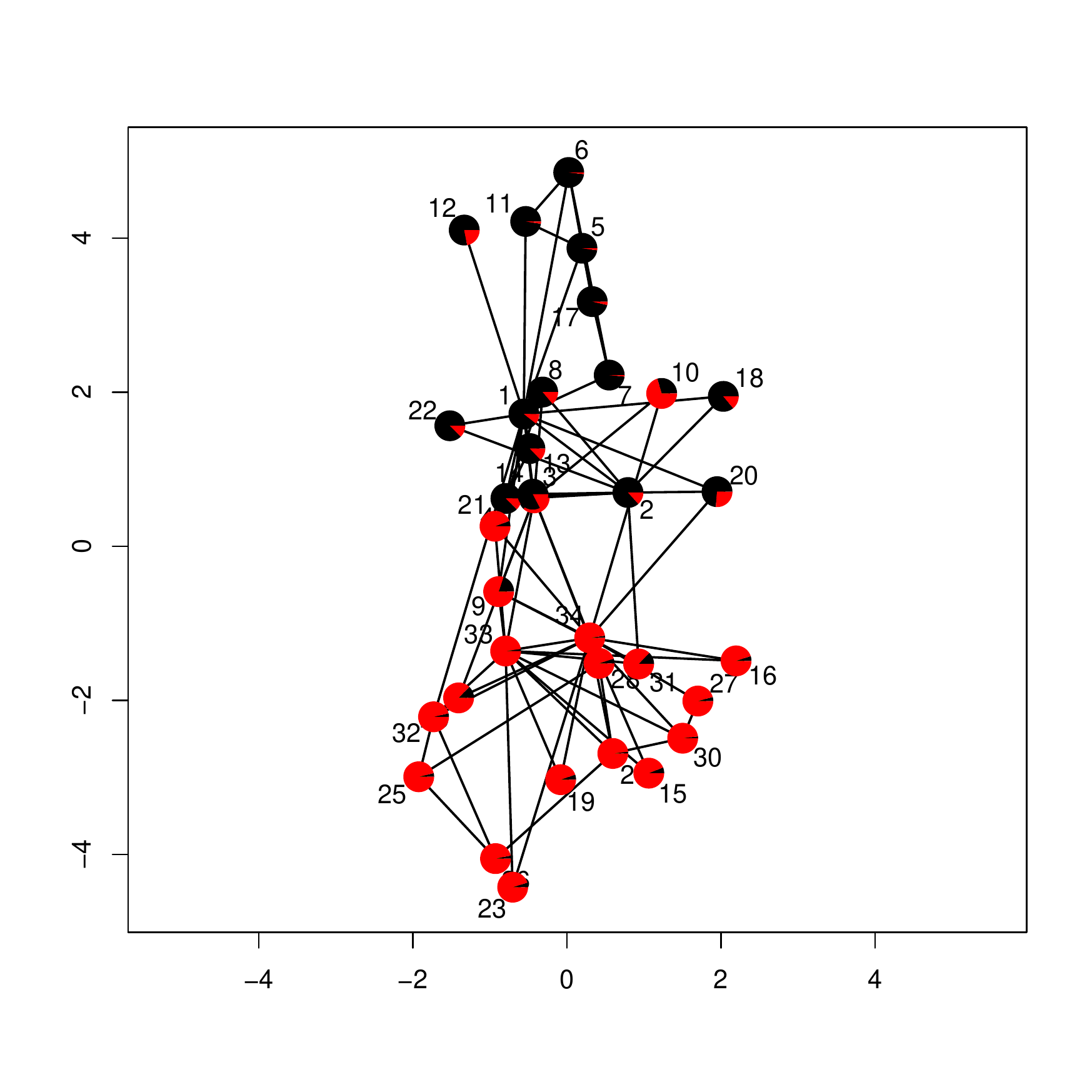} 
\includegraphics[width=7cm]{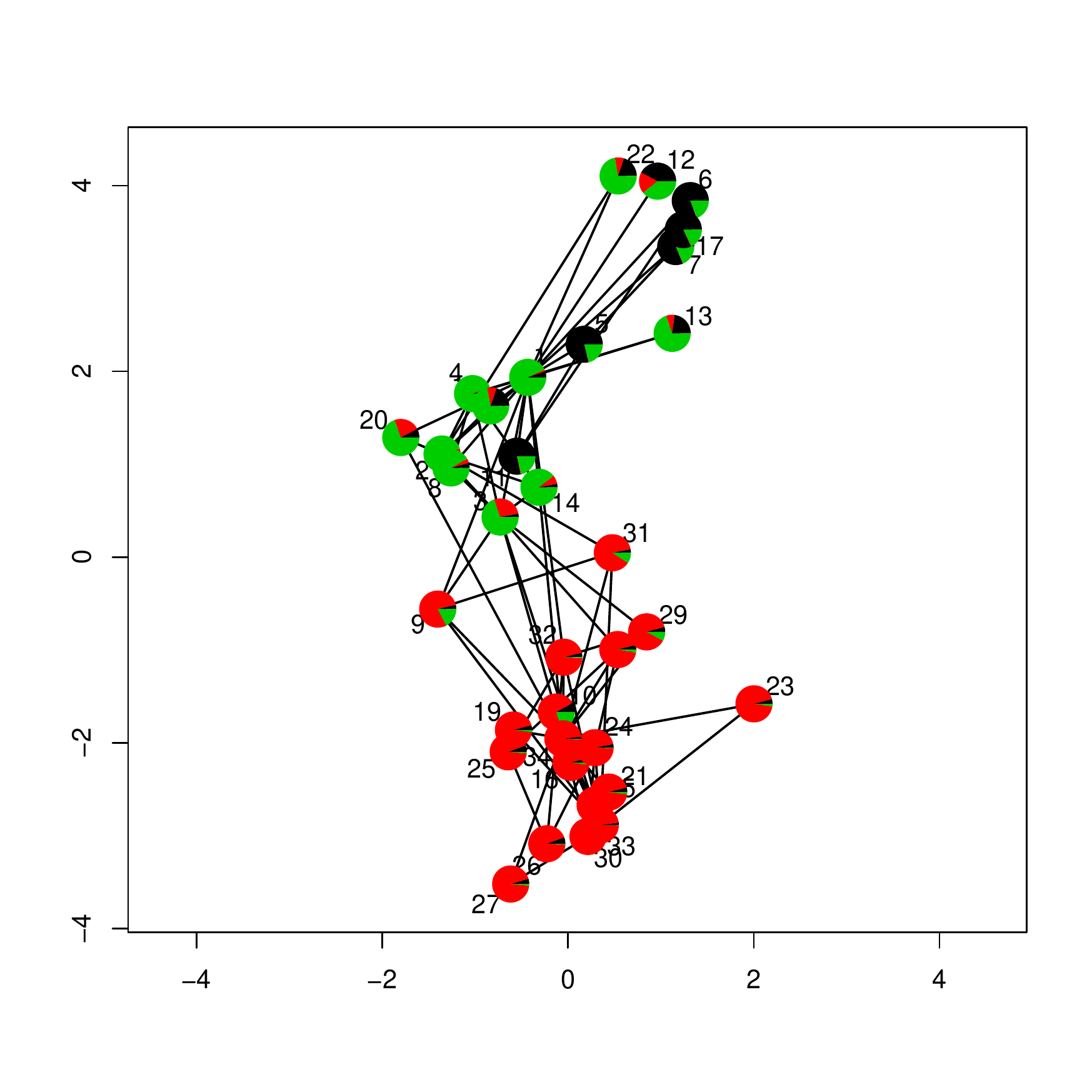} 
 \end{center}
 \caption{Zachary's karate club posterior mean actor positions using the collapsed sampler for the most probable  $2$ and $3$ group models with a pie chart depicting uncertainty of cluster memberships.}
\label{zmeanpiezac}
\end{figure}

Posterior mean actor positions for the \texttt{collapsed} sampler are shown in Figure \ref{zmeanpiezac}. 
There is good agreement between the actual club split and the clustering of our friendship network  for the $2$ group model.  
Interestingly, actor 9 is clustered with the coach Mr. Hi in our analysis
whose club he stayed in, due to the fact that he was only three weeks away from a test for his black belt (master status) when the split in the club
 occurred \cite{zachary77}. The posterior probability of membership was 0.8 to stay in the coach's group and 0.2 to go with the president.
Combining two clusters of the $3$ group model mirrors the true split as before. 

\begin{table}[htp]
\begin{center}
\begin{tabular}{| l|c|c|c|c|c|}
\hline
 & $G=1$ & $G=2$ & $G=3$ & $G=4$ & $G=5$ \\
\hline
	\texttt{collapsed} & 0.15 & 0.33 & \bf{0.40}  & 0.10  & 0.02  \\
  \texttt{latentnet} BIC & 537.46 & \bf{510.10} & 510.96 & 519.50 & 522.75 \\ 
  \texttt{VBLPCM} BIC & 1243.70  & 1130.09   & 1119.62  & 1108.39 & \bf{1104.89}\\ 
  \hline
\end{tabular}
\caption{Estimated posterior distribution of $G$ from our \texttt{collapsed} sampler, and approximate BIC using \texttt{latentnet} and \texttt{VBLPCM} for the karate network. Favoured model is indicated in bold font.}
\label{tab:pallzac}
\end{center}
\end{table}

The results are qualitatively similar for the \texttt{collapsed }and the \texttt{latentnet} $G=2$ group models. 
The \texttt{VBLPCM} algorithm chose a $5$ component model with practically no uncertainty in cluster membership. As a comparison of run times, fixing $G=2$ and running \texttt{latentnet} for $10^5$ burn-in iterations and a subsequent $10^6$ iterations storing every $100$th took 150 seconds. A run of the \texttt{collapsed} sampler for the same number of iterations took 249 seconds. We note however that the \texttt{collapsed} sampler output provides information on the most probable model indexed by $G$ and does more work in each iteration in order to do so. The \texttt{VBLPCM} fixing $G=2$ was the fastest taking 1.5 seconds. All times reported refer to computations on a single core of a 2.1GHz Intel Core i7 quad core processor.

\subsection{Dolphin Network}

The dolphin network studied by Lusseau \textit{et al} \citeyear{lusseauetal03} represents social associations between $62$ dolphins living off Doubtful Sound in New Zealand. It is an undirected graph with $159$ ties.

Proposal variances for the Metropolis-Hastings moves were $\sigma^2_{\x}=3$ and $\sigma^2_{\beta}=0.2$ for the latent actor positions and for the intercept respectively. Acceptance rates for the eject and absorb moves for this example were roughly 0.3\%. Higher rates were observed for the other examples. 

\begin{figure}[htp]
 \begin{center}
\[
\begin{array}{cc}
 \includegraphics[width=7.2cm]{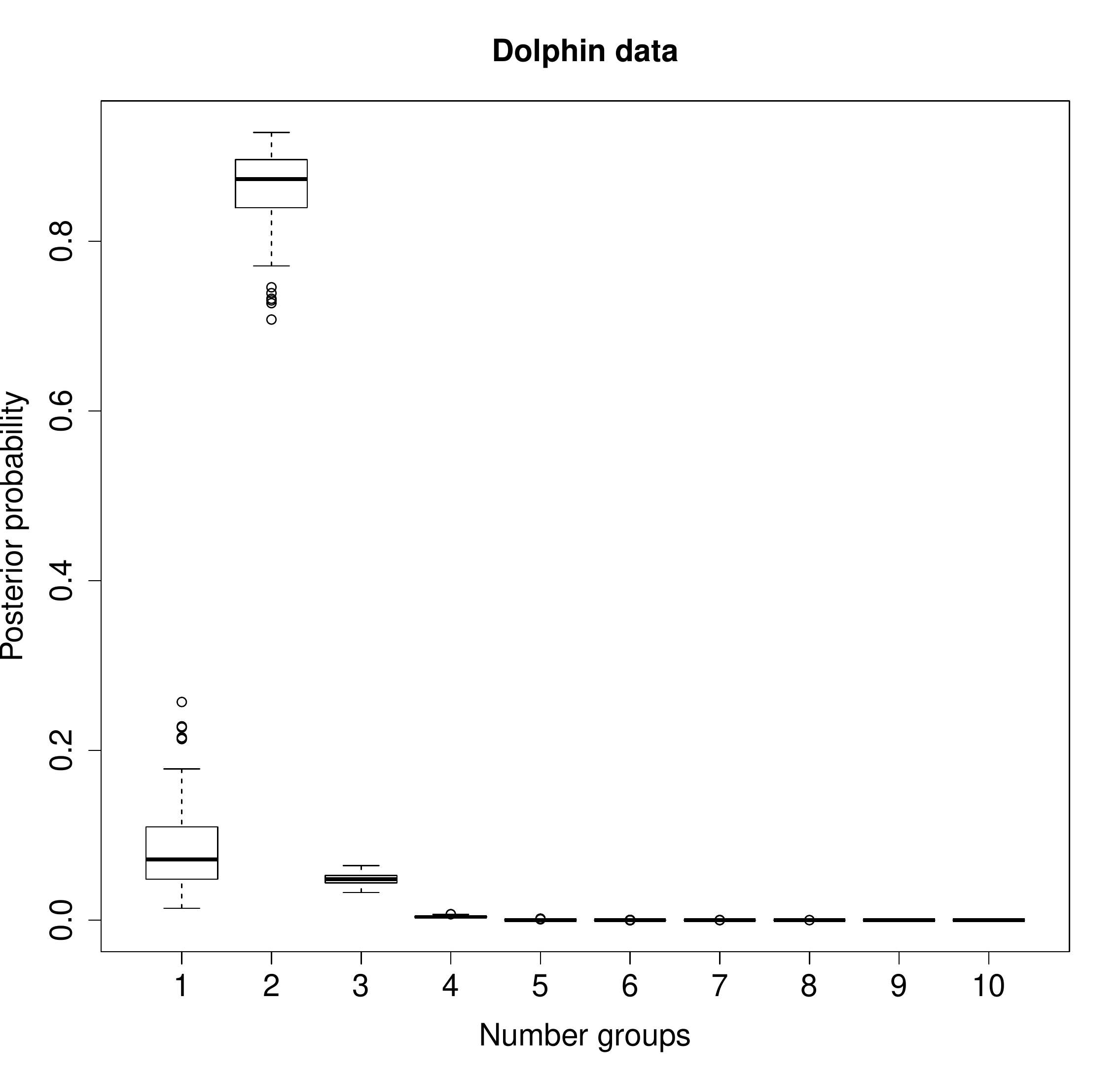}
 & \includegraphics[width=8cm]{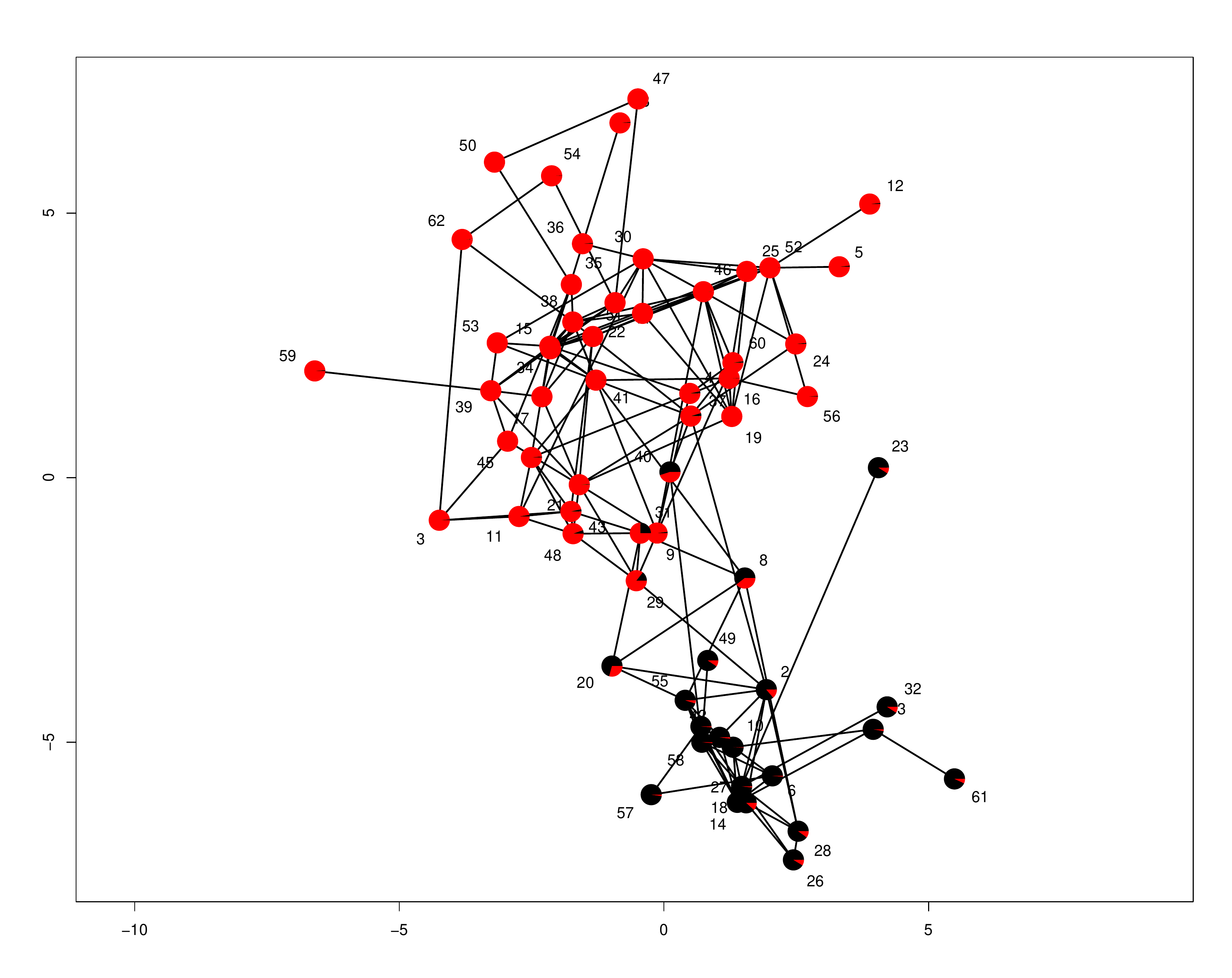} 
\end{array}
\]
 \end{center}
 \caption{ Dolphin network; left: barplots showing the posterior probability mass function of $G$ over 100 runs of the sampler; right: posterior mean actor positions using the collapsed sampler for the most probable $2$ group model with a pie chart depicting uncertainty of cluster memberships.}
\label{fig:zmeanpiedolph}
\end{figure}

A $2$ group model had highest posterior mass from our sampler output and approximate BIC using \texttt{latentnet}. Posterior model probabilities based on 100 runs of the \texttt{collapsed} sampler are displayed in Figure~\ref{fig:zmeanpiedolph} (left)
with the posterior from our sampler and inferred BIC approximations to the approximated model evidence using \texttt{latentnet} and \texttt{VBLPCM} given in Table~\ref{tab:palldolph}.

\begin{figure}[htp]
 \centering
\includegraphics[width=7cm]{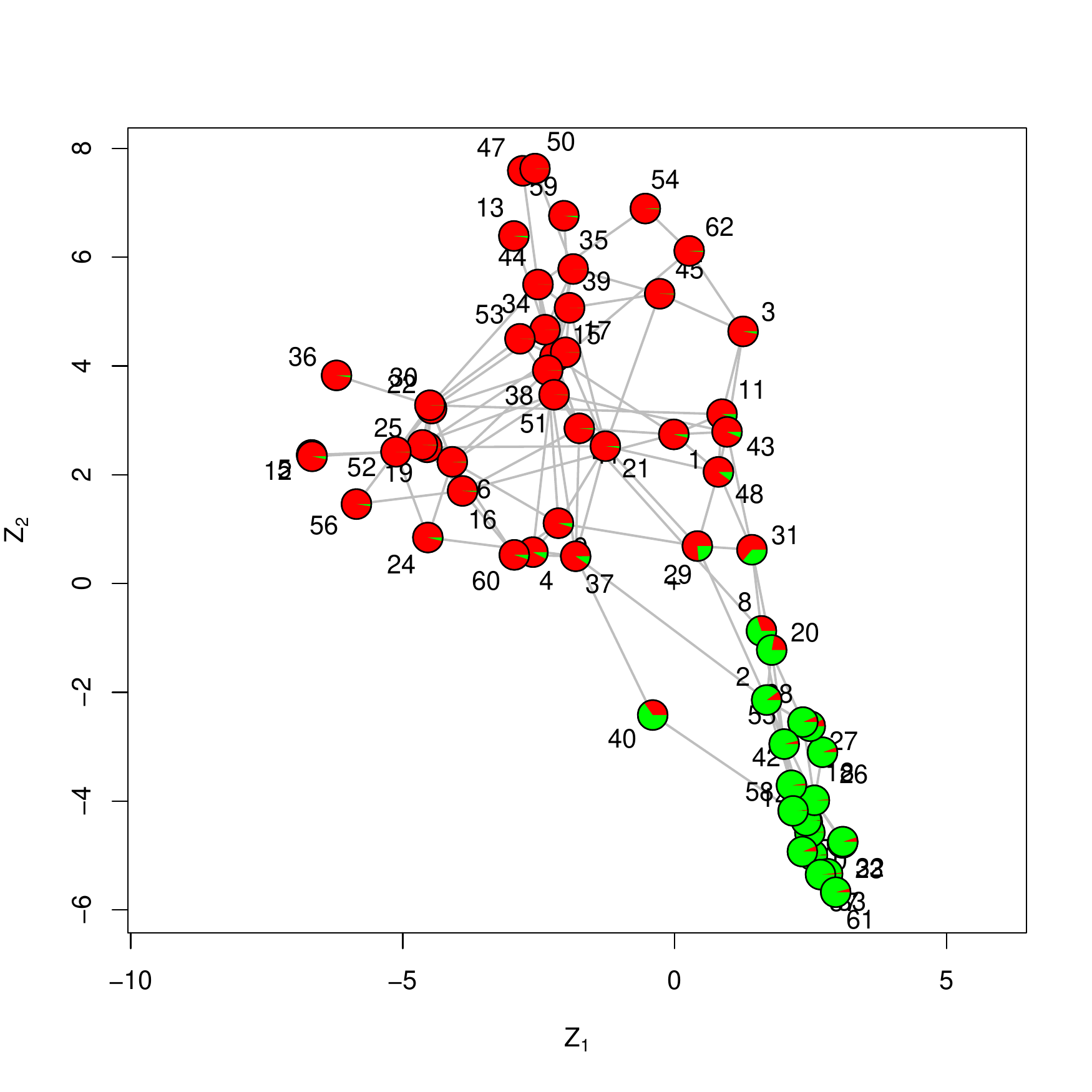}
\includegraphics[width=7cm]{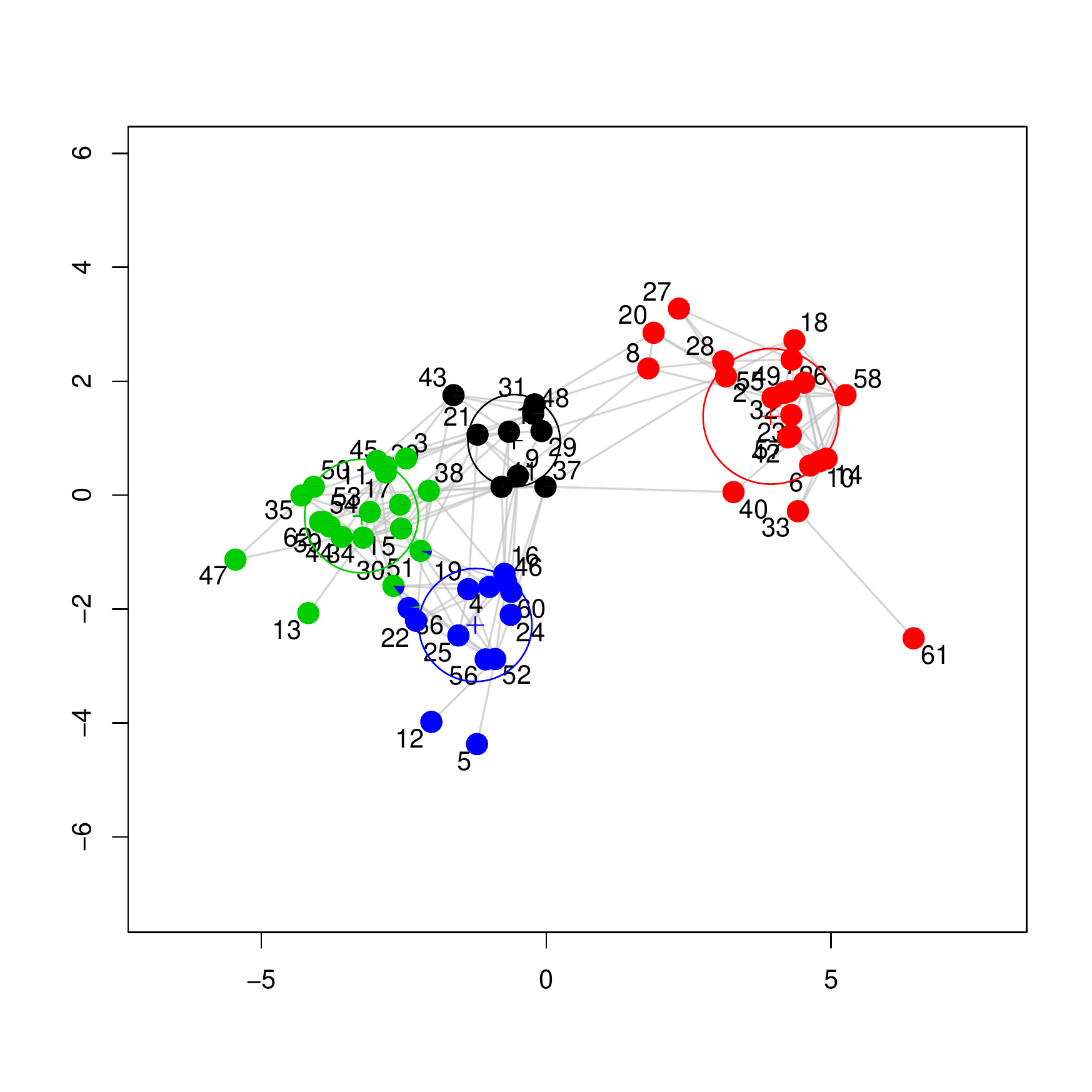}
\caption{Dolphin network analysis using other approaches; left: posterior mean \texttt{latentnet} positions and uncertain clustering for the $2$ component model of the dolphin network; right: \texttt{VBLPCM} estimates of latent positions for the $4$ component model for the dolphin network.}
\label{fig:lnetdolph}
\end{figure}

Posterior mean actor positions inferred using one run of the \texttt{collapsed} sampler are displayed in Figure~\ref{fig:zmeanpiedolph} (right). 
From Figure~\ref{fig:lnetdolph} (left),  good agreement can be seen between inference using \texttt{latentnet} and the \texttt{collapsed} sampler choosing the $2$ group model with qualitatively similar estimates of the latent actor positions (modulo a rotation) as well as allocations. Results inferred by \texttt{VBLPCM} differed (Figure~\ref{fig:lnetdolph}, right), favouring the $4$ group model with very little uncertainty in group membership. 

\begin{table}[htp]
\begin{center}
\begin{tabular}{| l|c|c|c|c|c|}
\hline
 & $G=1$ & $G=2$ & $G=3$ & $G=4$ & $G=5$ \\
\hline
	\texttt{collapsed} & 0.04 & \bf{0.91} & 0.05  &  0.00  & 0.00  \\
  \texttt{latentnet} BIC & 1176.53 & \bf{1149.60} & 1158.36 & 1168.80 & 1181.05  \\ 
  \texttt{VBLPCM} BIC & 2911.28  &  2488.08  & 2464.02  & \bf{2362.88} & 2537.20 \\ 
  \hline
\end{tabular}
\caption{Estimated posterior distribution of $G$ from our \texttt{collapsed} sampler, and approximate BIC using \texttt{latentnet} and \texttt{VBLPCM} for the Dolphin network. Favoured model is indicated in bold font.}
\label{tab:palldolph}
\end{center}
\end{table}

\section{Discussion} \label{discussion}
A novel approach to model selection for the latent position cluster model for social networks has been presented. 
Use of conjugate priors allows most of the clustering parameters to be marginalized out from the model analytically and provides a fixed dimensional parameter space for trans-model inference. 
This admits joint inference on the number of clusters in the network and latent positions simultaneously. It avoids multiple approximations used by~\citeasnoun{handcocketal07} to evaluate the model evidence,
while improving computational efficiency compared with standard methods.
Parallelization is possible for the likelihood (Section~\ref{sec:betaupdate}), but not exploited in this paper and could give further
decreases in compute time.

The simulation study in Section~\ref{sec:simulated} showed that our approach gave sensible results over networks with varying levels of cluster separation when the true number of clusters is known. Our \texttt{collapsed} sampler was then demonstrated three real data examples with comparison to current state-of-the-art methods for the LPCM. Similar results were found between our methods and sampling the full posterior for separate models using \texttt{latentnet}~\cite{Kriv:Hand13}. Substantial uncertainty in the number of clusters and cluster membership was evident. The model and sampler proposed thus provided a way to quantify this uncertainty in the model structure indexed by $G$ on a natural probability scale. 
Software implementing the methods in this paper is available from \texttt{https://www.scss.tcd.ie/Jason.Wyse}.

\paragraph*{Acknowledgements:} Caitr\'{i}ona Ryan and Nial Friel's research was supported by a Science Foundation Ireland
Research Frontiers Program grant, 09/RFP/MTH2199. This research was also supported in part by a research grant from Science 
Foundation Ireland (SFI) under Grant Number SFI/12/RC/2289. Jason Wyse's research was supported in part through the STATICA project, a 
Principal Investigator program of Science Foundation Ireland, 08/IN.1/I1879.

\appendix

\section{Simulation study design} \label{sec:appendix}

Fixing $\beta=0$ in the LPCM model, the probability of a link between two actors with latent positions $\x$ and $\x'$ is
$p( \x, \x') = 1/(1 + \mathrm{e}^{-\eta})$, where $\eta = - ||\x - \x'||$. We assume that $\x, \x'$ are independent with $\x \sim \mathcal{N}(\bfmu, 1/\tau\mathbf{I})$
and $\x' \sim \mathcal{N}( s \bfmu, 1/\tau\mathbf{I})$. Note here that $s=1$ corresponds to $\x$ and $\x'$ being drawn from a $\mathcal{N}(\bfmu, 1/\tau \mathbf{I})$, so that their (owning) actors in the network are in the same cluster with centre $\bfmu$. 
On the other hand, $s=-1$ corresponds to two actors belonging to opposite clusters (cluster 1 and cluster 2 respectively). The probability of a link between two arbitrary actors in cluster 1 can be written using
\begin{equation}
\mathrm{E}_s(p) = \int \int p( \x, \x')\, \mathcal{N}(\x; \bfmu, 1/\tau  \mathbf{I}) \, \mathcal{N}( \x'; s\bfmu, 1/\tau\mathbf{I})\, \diff \x \,\diff \x'. \label{eq:ex_p}
\end{equation}
and taking $s=1$. Taking $s=-1$ in (\ref{eq:ex_p}) gives the probability of a link between two arbitrary actors in clusters 1 and 2 respectively. 
As $\bfmu_2$ is the reflection of $\bfmu_1$ through the origin, the probability of a link between two arbitrary actors when $\x$ and $\x'$ are both from $\mathcal{N}(\bfmu_2, 1/\tau \mathbf{I})$ will be equal to $\mathrm{E}_1(p)$. 
Using (\ref{eq:ex_p}) it can be shown that the conditional probability of a tie given two arbitrary actors come from the same cluster is given by $0.5\times \mathrm{E}_1(p)$. The conditional probability of a tie given two arbitrary actors come from different clusters is given by $0.5\times \mathrm{E}_{-1}(p)$.
The difficulty of clustering the network will be determined by how probable actors are to have ties to those in the same cluster as compared to having ties to actors in the other cluster. We use the ratio of the probabilities of within cluster to between cluster ties
\[
r = 0.5 \times \mathrm{E}_{1}(p_{ij}) / 0.5 \times \mathrm{E}_{-1}(p_{ij})
\]
as a measure of the difficulty of the clustering task. A value of $r=1$ implies no notable difference in linking propensity whether two actors are in the same or opposite clusters. As $r$ increases, the actors should be clustered more easily. For a specified value of $r$ we determine approximate values for $\mu$ and $\tau$ which will produce such a network. To do this we take a grid of 20 equally spaced values of $\mu \in [0.1,2]$ and $\tau \in [1,20]$ and for each $(\mu,\tau)$ pair simulate $N=10,000$ latent positions $\x_t,\x'_t, t=1,\dots,N$. We then estimate (\ref{eq:ex_p}) for $s=-1,1$ by
\[
\mathrm{E}_s(p) \approx \frac{1}{N} \sum_{t=1}^N p( \x_t, \x'_t)\, \mathcal{N}(\x_t; \bfmu, 1/\tau  \mathbf{I}) \, \mathcal{N}( \x'_t; s\bfmu, 1/\tau\mathbf{I}).
\]
This allows us to produce a lookup table for $r$. For a specified $r$ we find the $(\mu,\tau)$ pair in the table which produce the closest match to $r$.

\bibliography{bibliography_RyanWyseFriel}

\end{document}